\documentclass[article,onecolumn,preprint,nofootinbib,prd]{revtex4}
\pdfoutput=1
\usepackage{graphicx,amsfonts,color,comment,amsmath,float}
\usepackage{amssymb}	
\usepackage{scrextend}

\usepackage{mathrsfs,amssymb}  
\usepackage{cancel}
\usepackage[normalem]{ulem}

\newcommand{\be}{\begin{equation}}
\newcommand{\ee}{\end{equation}}
\newcommand{\bea}{\begin{eqnarray}}
\newcommand{\eea}{\end{eqnarray}}

\usepackage{color}
\usepackage{graphicx}  
\usepackage{xcolor}
\begin{document}

\title{A Cusp-Core like challenge for Modified Newtonian Dynamics}

\date{\today} \author{Mikkel H. Eriksen}\email{mheriksen@cp3.sdu.dk}
\date{\today} \author{Mads T. Frandsen}\email{frandsen@cp3.sdu.dk}
\date{\today} \author{Mogens H. From}\email{from@cp3.sdu.dk}
\affiliation{$CP^3$-Origins, University of Southern Denmark, Campusvej 55, DK-5230 Odense M, Denmark}
\date{\today}

\begin{abstract}
We show that Modified Newtonian Dynamics (MOND) predict distinct galactic acceleration curve geometries in $g2$-space --- 
the space of total observed centripetal accelerations  $g_{\rm tot}$ vs the inferred Newtonian acceleration from baryonic matter $g_{\rm N}$ --- and corresponding rotation speed curves: 
MOND modified gravity predicts cored geometries for isolated galaxies while MOND modified inertia yields neutral geometries, ie. neither cuspy or cored, based on a cusp-core  classification of galaxy rotation curve geometry in $g2$-space  --- rather than on inferred DM density profiles. 

The classification can be applied both to DM and modified gravity models as well as data 
and implies a {\it cusp-core} challenge for MOND from observations, for example of cuspy galaxies, which is different from the so-called cusp-core problem of dark matter (DM).
We illustrate this challenge by a number of cuspy and also cored galaxies from the SPARC rotation curve database, which deviate significantly from the MOND modified gravity and MOND modified inertia predictions.

\end{abstract}
\preprint{CP³-Origins-2019-24 DNRF90}

%\pacs{}
\maketitle

\section{introduction}
\noindent 
The missing mass problem in astrophysical systems from galaxies and galaxy clusters to the CMB is well established. Gravitational potentials are observed to be deeper than predicted from the visible matter distributions in Newtonian gravity.
Early observations of this phenomenon include the velocity dispersion of galaxies in clusters \cite{Zwicky:1933gu} and galactic rotation curves \cite{Rubin:1970zza,Rubin:1980zd,Bosma:1981zz}.
Both particle Dark Matter \cite{Lee:1977ua,Steigman:1984ac} (DM) and Modified Newtonian Dynamics  (MOND) \cite{Milgrom:1983ca} were proposed as explanations of these observations. In MOND the acceleration of test particles is modified, with respect to the Newtonian prediction, below a characteristic acceleration scale $g_0\sim 10^{-10} m/s^2$. 
to yield asymptotically constant speeds in rotation curves at large radii and low accelerations \cite{Sanders:2002pf,Gentile:2010xt,McGaugh:2016leg,Lelli:2017vgz,Li:2018tdo} as observed. It also provides a correlation of this asymptotic speed with the total baryonic mass in the galaxy, i.e. the baryonic Tully-Fisher relation \cite{Tully:1977fu,McGaugh:2000sr}.
However, it has been argued that MOND cannot account for the entire missing mass observed in galaxy clusters
\cite{Sanders:2002ue} and today the more recent observations of merging clusters \cite{Clowe:2006eq} and the measurements of the cosmic microwave background \cite{Skordis:2005xk,Dodelson:2006zt,Dodelson:2011qv} is considered by many as a challenge for MOND. For a review of MOND and observations, see \cite{Famaey:2011kh}. Despite these known challenges, it is of obvious interest to investigate in detail the predictions of MOND for rotation curves beyond the asymptotic velocities at large radii.

Recently the entire sample of Galaxies in the SPARC data base was compared to a MOND modified inertia model in and it was found that the fit residuals were gaussian and of the expected size \cite{McGaugh:2016leg,Lelli:2017vgz,Li:2018tdo} . However in \cite{Frandsen:2018ftj} it was shown that data at small radii deviate significantly from the MOND modified inertia fit. Since the sample of data at small radii is a few hundred points compared to the few thousand points at large radii the discrepancy is only apparent in the residuals if these data are treated separately (see also Fig.~\ref{Fig: residuals} below) or if galaxies are considered individually \cite{Petersen:2017klw}  (see also Fig.~\ref{Fig: Galaxiescorecusp} below). Investigating data points at small radii separately is well motivated as this is where the predictions of different models of the missing mass problem deviate  significantly. In particular MOND modified gravity and MOND modified inertia models only deviate in their predictions at small radii, and in a definite manner, as we show here.

\bigskip
Early simulations of structure formation with cold collisionless DM particles and no baryons found universal cuspy NFW-like DM density profiles in halos ranging from dwarf galaxies to galaxy clusters \cite{Navarro:1995iw}. This profile fits rotation curve data at large radii in galaxies and clusters, but not in all cases at small radii. 
The inferred DM densities from some observed clusters and gas-rich halo dominated dwarf spirals is less steep, i.e. more cored, than the NFW profile in the inner regions \cite{Flores:1994gz}. This has become known as the cusp-core problem for DM. More recent DM only simulations find some systematic departures from the NFW profile and some diversity in resulting rotation curves \cite{Navarro:2008kc}. 
But still these DM only simulations show little variation in rotation curve profiles with the same asymptotic maximal rotational velocity \cite{Oman:2015xda} while observed rotation curves of dwarf galaxies do show such a variation. This has been termed the diversity problem. 
Whether the cusp-core problem or the diversity problem is a problem of DM, or rather of simulations with limited resolution and without inclusion of baryons remains debated, as simulations with baryonic feedback included do find cored profiles \cite{Read:2004xc,Teyssier:2012ie,DiCintio:2013qxa}.

\bigskip
In this paper we identify a different cusp-core challenge pertaining to MOND which is essentially the opposite of that for DM. To do so we first provide a definition and classification of cuspy and cored galaxies based on 
acceleration curve geometry in $g2$-space following \cite{Frandsen:2018ftj} --- i.e. 
the space of total centripetal accelerations  $g_{\rm obs}$ vs the Newtonian centripetal acceleration from baryonic matter $g_{\rm bar}$ --- rather than on inferred DM density profiles. 
This classification is directly applicable to both MOND and DM models. We show that MOND modified gravity models, in the Bekenstein-Milgrom formulation, lead to cored acceleration curve geometries in $g2$-space and corresponding rotation speed curves --- a consequence of the so-called solenoidal acceleration field in these models. We first illustrate this using analytical approximations \cite{Brada:1994pk} and by investigating the curl of the solenoidal field, before we explicitly solve the MOND modified Poisson equation using the {\tt N-Mody} code \cite{Ciotti:2005bi}. In contrast MOND modified inertia curves provide a definition of 'neutral' curves, neither cuspy nor cored, as benchmark.

A way to test MOND modified gravity is therefore to look for cuspy galaxies as well as cored galaxies which deviate from the specific cored geometries predicted by MOND modified gravity models.  
We therefore compare the SPARC rotation curve data of selected galaxies with predictions from MOND modified gravity models,
and find that also the observed geometry of the cored galaxies is different from that predicted for MOND. 
This extends our previous analyses of MOND modified inertia models \cite{Petersen:2017klw,Frandsen:2018ftj}.

\bigskip
The paper is organized as follows. 
In Section~\ref{Sec:Classification} we present our definition and classification of cuspy, neutral, and cored galaxies based on rotation curve geometry in $g2$-space. 
In Secton~\ref{Sec: geometries} we discuss MOND modified gravity and MOND modified inertia models. We show that MOND modified gravity yields cored rotation curve geometries, which tend to neutral geometries for spherical mass distributions. We show this using general properties of the MOND solenoidal field, analytic approximations and full numerical solutions using the {\tt N-Mody} code \cite{Ciotti:2005bi}. Instead MOND modified inertia leads to neutral rotation curve geometries universally, independent of the matter distribution.

\section{Classification of cuspy and cored geometries in $g2$-space }
\label{Sec:Classification}
\begin{figure}[htp!]
	\centering
	\includegraphics[width=0.45\textwidth]{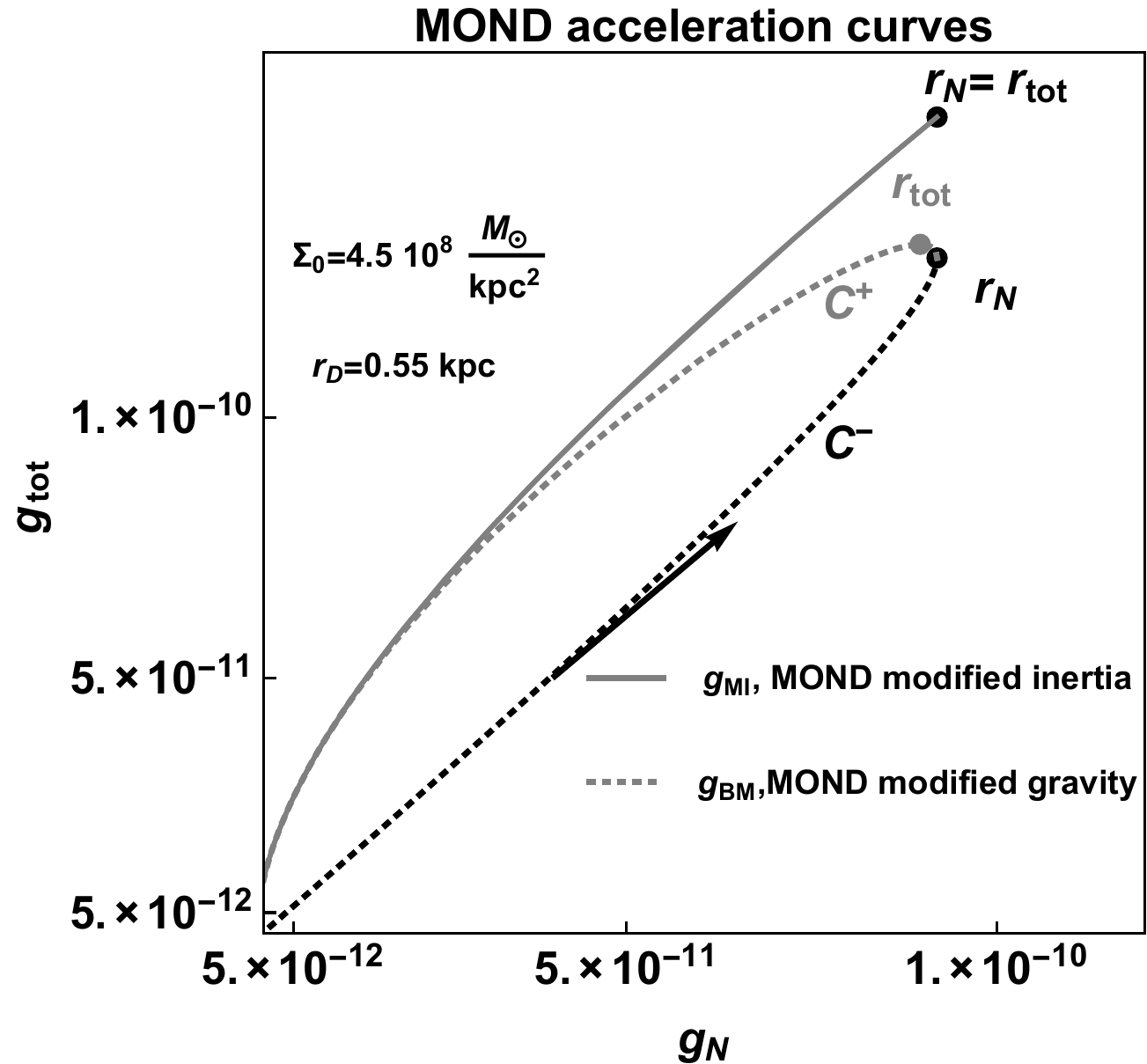}		
	\includegraphics[width=0.4\textwidth]{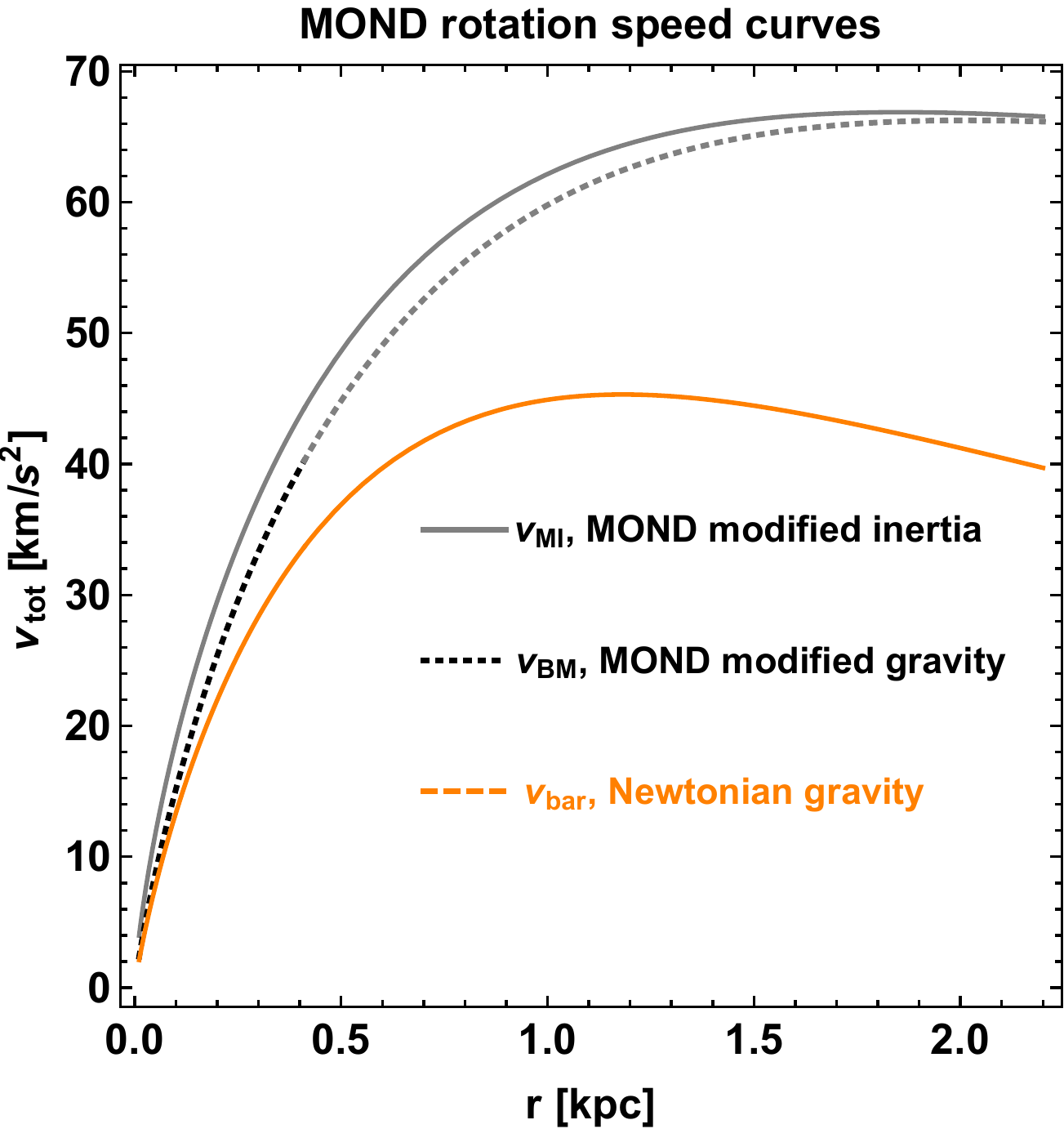}
	\caption
	{Left panel: Centripetal acceleration curves in $g2$-space with the quantities $r_{\rm tot}, r_N$ and $\mathcal{C^{\pm}}$ shown, used for classification in table~\ref{Table:geometries}. The grey solid line shows a MOND modified inertia curve with the radius of maximum baryonic acceleration and maximum total acceleration $r_N=r_{\rm tot}$ indicated. Also the curve segments $\mathcal{C^{+}}$ and $\mathcal{C^{-}}$ coincide so the curve area is $\mathcal{A}(\mathcal{C})=0$. The grey dotted and and black dotted curves show the  $\mathcal{C^{\pm}}$ curve segments of a MOND modified gravity curve, using the Brada-Milgrom approximation in Eq.~\eqref{Eq:MONDgrav}, with $r_{\rm tot}>r_N$ and $\mathcal{A}(\mathcal{C})>0$. The baryonic matter is for both  an infinitely thin exponential disk $\Sigma(r)=\Sigma_0 e^{-r/r_d}$. The arrow indicates the direction of increasing radius along the curve. Right panel: The corresponding rotation curves}
	\label{Fig:geometries}
\end{figure}

We begin by reviewing some geometric characteristics of galactic centripetal acceleration curves which we refer to as $g2$-space curves --- 
the space of total predicted\footnote{As in \cite{Frandsen:2018ftj} we use $g_{\rm tot}$ for total predicted model accelerations and $g_{\rm obs}$ for the total observed. Similarly we use $g_{\rm N}$ for the Newtonian accelerations from baryons in a given model and $g_{\rm bar}$ for the same inferred quantitiy from data.} centripetal accelerations  $g_{\rm tot}$ vs the Newtonian acceleration from baryons $g_{\rm N}$ --- following \cite{Frandsen:2018ftj}. 
Examples of MOND modified gravity and MOND modified inertia curves in $g2$-space are shown in Fig.~\ref{Fig:geometries} left panel and in Fig.~\ref{Fig:MONDcurves} left panel. 
Examples of $g2$-space data curves --- the space of total {\it observed} centripetal accelerations  $g_{\rm obs}$ vs the inferred Newtonian acceleration from baryons $g_{\rm bar}$ --- from the {\tt SPARC} database are shown in Fig.~\ref{Fig: Galaxiescorecusp} top panels.

We first define  
the radii, $r_{\rm tot}$ and $r_{\rm N}$ as the locations at which the centripetal accelerations $g_{\rm tot}$  and $g_{\rm N}$ are maximum:
\begin{equation}
g_{\rm N}(r_{\rm N}) = {\rm max} \{ g_{\rm N}(r) \} , \quad g_{\rm tot}(r_{\rm tot}) = {\rm max} \{ g_{\rm tot}(r) \}  \ . 
\label{Eq:refradii}
\end{equation}
These radii are indicated in the left panel of Fig.~\ref{Fig:geometries} and from these radii we define the acceleration ratios
\begin{align}
\hat{g}_{\rm tot}(r) \equiv g_{\rm tot}(r)/g_{\rm tot}(r_{\rm N}) , \quad \hat{g}_{\rm N}(r) \equiv g_{\rm N}(r)/g_{\rm N}(r_{\rm N}) .
\label{Eq:normg2-space}
\end{align}
When we later study {\tt SPARC} data, the equivalent data ratios $\hat{g}_{\rm bar}(r_{j,G})$ and  $\hat{g}_{\rm obs}(r_{j,G})$, of the measurements at the $j$th radius point in the Galaxy $G$ will be free of relevant systematic uncertainties \cite{Frandsen:2018ftj} as we return to below.

We will classify galaxies as cuspy if $r_{\rm tot}<r_{\rm N}$, neutral if  $r_{\rm tot}=r_{\rm N}$ and cored if $r_{\rm tot}>r_{\rm N}$. More generally we are interested in the relative location of the entire curve segment  ${C}^{+}$ at large radii and that at small radii ${C}^{-}$ defined with respect to some reference radii $r_*$ \cite{Petersen:2017klw}. In this study we take $r_N$ and so define
\begin{equation}
\mathcal{C}^{+}= \{ (g_{\rm N}(r),  g_{\rm tot}(r)) ; r >  r_{\rm N}\}  , \quad \mathcal{C}^{-}= \{ (g_{\rm N}(r),  g_{\rm tot}(r)) ; r <  r_{\rm N}\}
\label{Eq:refcurves}
\end{equation}
Galaxies are cuspy if $\mathcal{C}^{+}<\mathcal{C}^{-}$, in the sense that the former curve segment lies above the latter, they are neutral if  $\mathcal{C}^{+}=\mathcal{C}^{-}$ and cored if  $\mathcal{C}^{+}>\mathcal{C}^{-}$.
Finally we can classify $g2$-space model curves according to whether they are open curves as a NFW profile DM model or closed curves with $(g_{\rm N}(\infty),  g_{\rm tot}(\infty))= (g_{\rm N}(0),  g_{\rm tot}(0))=(0,0)$. If they are closed we can further classify the curves according to the signed curve area $\mathcal{A}(\mathcal{C})$ with curves running counterclockwise, as parameterized by the radius $r$ running from $r=0$ to $r=\infty$, defined to have positive area.
 From the quantites $r_{N,tot}$,  ${C}^{\pm}$ and $\mathcal{A}(\mathcal{C})$ we define cuspy, neutral and cored geometries as in table~\ref{Table:geometries}.
 \begin{table}[htp!]
\centering
\begin{tabular}{ |p{2.5cm}||p{3cm}|p{5cm}|p{3cm}|p{4cm}| }
 \hline
Models& Reference radii & Accelerations & Curve segments& Signed Curve Area\\
 \hline
Cuspy &   $r_{\rm tot}<r_{\rm N}$ & $\hat{g}_{\rm tot}(r_{\rm tot})>1$& $\mathcal{C}^+<\mathcal{C}^-$&$\mathcal{A}(\mathcal{C})<0$ or open curve \\
Neutral &   $r_{\rm tot}=r_{\rm N}$ & $\hat{g}_{\rm tot}(r_{\rm tot})=1$& $\mathcal{C}^+=\mathcal{C}^-$&$\mathcal{A}(\mathcal{C})=0$\\
Cored &   $r_{\rm tot}>r_{\rm N}$  & $\hat{g}_{\rm tot}(r_{\rm tot})>1,$ &$\mathcal{C}^+>\mathcal{C}^-$&$\mathcal{A}(\mathcal{C})>0$\\
 \hline
\hline
Data&  & & & \\
 \hline
Cuspy &   $r_{\rm obs,G}<r_{\rm bar,G}$ & $\hat{g}_{\rm obs,G}(r_{\rm obs,G})>1$& $\mathcal{C}^+<\mathcal{C}^-$&\\
Neutral &   $r_{\rm obs,G}=r_{\rm bar,G}$ & $\hat{g}_{\rm obs,G}(r_{\rm obs,G})=1$& $\mathcal{C}^+=\mathcal{C}^-$&\\
Cored &   $r_{\rm obs,G}>r_{\rm bar,G}$  & $\hat{g}_{\rm obs,G}(r_{\rm obs,G})>1$ &$\mathcal{C}^+>\mathcal{C}^-$&\\
 \hline
\end{tabular}
\caption{Global characteristics of cuspy, neutral and cored geometries of rotation acceleration curves for models (top three rows) and for data (bottom three rows)in $g2$-space. The characterization applies to modified gravity and DM models alike. The reference radii $r_{\rm tot}$ and $r_{\rm N}$ are the radii of maximum total acceleration and maximum baryonic acceleration, as defined in Eq.~\eqref{Eq:refradii}. The reference radii  $r_{\rm obs}$ and $r_{\rm bar}$ are the analogues in data.  The curve segments $\mathcal{C}^{\pm}$ are defined in Eq.~\eqref{Eq:refcurves}.
}
\label{Table:geometries}
\end{table}

Our definition is more general than that normally used for DM profiles, but a DM model with NFW like profile is cuspy also according to our definition while that of DM with a quasi-isothermal profile is cored as illustrated in \cite{Frandsen:2018ftj}. From the right panel of Fig.~\ref{Fig:geometries} it is also seen that the MOND modified gravity rotation speed curve   indeed has a more shallow approach to zero radius relative to the MOND modified inertia rotation speed curve. This would correspond to a more cored density profile in the former case if it arose from DM. It is also seen that while the difference in $g2$-space curves is very significant the effect is modest at the level of the rotation speed curve.

\section{MOND Models and their $g2$-space geometries}
\label{Sec: geometries} 
In this section we show that MOND modified gravity leads to cored geometries with $r_{\rm tot}>r_{\rm N}$, $\mathcal{C}^+>\mathcal{C}^-$ and $\mathcal{A}(\mathcal{C})>0$ for isolated galaxies with axisymmetric mass distributions. The resulting rotation speed curves are more shallow than MOND modified inertia which universally leads to the geometries which we here term neutral. As were already discussed in \cite{Frandsen:2018ftj} they are characterized by $r_{\rm tot}=r_{\rm N}$, $\mathcal{C}^+=\mathcal{C}^-$ and $\mathcal{A}(\mathcal{C})=0$. Examples of MOND modified gravity (black and coloured dotted lines) and MOND modified inertia (grey solid lines) curves in $g2$-space are displayed in the left panels of Fig.~\ref{Fig:geometries} and in Fig.~\ref{Fig:MONDcurves}.

The cored MOND modified gravity geometries reduce to the neutral MOND modified inertia geometries for spherical mass distributions. To show the cored geometries of MOND modified gravity we start from analytical approximations and the simpler Quasilinear MOND modified gravity models \cite{Milgrom:2009ee} and finally solve the MOND modified gravity curves explicitly using the {\tt N-MODY} code \cite{Ciotti:2005bi}.

{\bf MOND modified gravity: }  In the Bekenstein-Milgrom formulation of MOND modified gravity models \cite{Bekenstein:1984tv} the total centripetal acceleration in a galaxy is determined via a modified Poisson equation for the MOND potential field $\psi$
\begin{equation}
\vec{\nabla} \cdotp (\mu(\frac{| \vec{\nabla}\psi |}{g_0}) \vec{\nabla} \psi  ) = 4 \pi G \rho . 
\label{Eq:MONDgravstart}
\end{equation}
In this equation $- \vec{\nabla}\psi=\vec{g}_M$ is the MOND acceleration and $g_0\sim 10^{-10} \frac{m}{s^2}$ is a characteristic acceleration scale such that the {\it interpolation function} $\mu(x)$ smoothly interpolates between the Newtonian regime $\mu(x)\simeq 1$ for $x\gg 1$ and the deep mondian regime $\mu(x)\simeq x$ for $x\ll 1$ with  $x \mu(x)$ monotonic. 
The limiting behaviour of $\mu(x)$ for $x\gg 1$  is clearly required to recover Newtonian dynamics and the limiting behaviour in the deep Mondian limit leads to constant rotation curve speeds at large radii. The modified Poisson equation,  Eq.~\eqref{Eq:MONDgravstart}, is derived from a general extension of the Lagrange for Newtonian gravity under the assumption that the acceleration $\vec{g}_M$ arises from a single potential \cite{Bekenstein:1984tv}.

Using the Poisson equation for the Newtonian potential $\Phi$, $4 \pi G \rho=  \vec{\nabla}^2 \Phi  = -  \vec{\nabla} \cdot \vec{g}_{\rm N}$, where $\vec{g}_{\rm N}$ is the acceleration predicted by Newtonian dynamics, the modified Poisson equation may be rewritten in terms of accelerations as 
\begin{equation}
\mu(\frac{g_M}{g_0})\vec{g}_M   =  \vec{g}_{\rm N} +  \vec{S} = \vec{q} , \quad  \quad \vec{g}_{\rm M}=\nu(\frac {q}{g_0})\vec{q} , 
\label{MONDgrav} . 
\end{equation}
where the solenoidal field $\vec{S}$ has zero divergence $\nabla \cdot \vec{S}=0$.
 The inverse interpolation function $\nu(y)$ is defined such that $\nu(y)\equiv I^{-1}(y)/y$ with $I(x)=x \mu(x) =y$. 
A number of interpolation functions $\mu(x)$ and inverse interpolation functions $\nu(y)$ have been considered in the literature, e.g. \cite{Begeman:1991iy,Bekenstein:2004ne}.  
Here we consider two inverse interpolation functions: The $\nu_1(y)$ function used in the {\tt N-MODY} code and the $\nu_2(y)$ proposed in \cite{Milgrom:2007br,McGaugh:2008nc,Famaey:2011kh} and used to fit the SPARC galaxy data in \cite{McGaugh:2016leg,Lelli:2017vgz}: 
\begin{equation}
\nu(y)_1=\sqrt{\frac{1}{2}(1+\sqrt{1+\frac{4}{y^2}})}  , \quad \nu(y)_2=\frac{1}{1-e^{-\sqrt{y}}} . 
\label{Eq:interpolationfunctions}
\end{equation}
\normalsize
The corresponding $\mu_1$ function  is $\mu_1(x)= \frac{x}{\sqrt{1+x^2}}$ while $\mu_2$ is transcendental.

\bigskip
{\bf MOND modified inertia: }  For spherical matter distributions $\vec{S}=\vec{0}$ \cite{Bekenstein:1984tv}, and it follows that in this special case the Newtonian and MOND ($\vec{g}_{MI}$) accelerations are related as
\begin{equation}
\vec{g}_{\rm N}=\mu(\frac{g_{MI}}{g_0})\vec{g}_{MI}  ; \quad \vec{g}_{\rm MI}=\nu(\frac{g_N}{g_0})\vec{g}_N ,  
\label{Eq: MONDI}
\end{equation}
These relations with $\vec{S}=\vec{0}$ also hold in so-called MOND modified inertia models \cite{Milgrom:1983ca} for any matter distributions. We therefore denote the MOND acceleration as $\vec{g}_{MI}$ in this case. 
MOND modified inertia is often used for comparing MOND with rotation curve data irrespective of the distinction between MOND modified inertia and MOND modified gravity. 
However,  the acceleration $\vec{g}_{MI}$ is not in general derivable from a potential as $\vec{g}_{M}$ defined via Eq.~\eqref{Eq:MONDgravstart} is and 
%However 
as seen above, the $g2$-space geometry of the two are distinctly different.

Since the function $x \mu(x)$ is monotonic, the function $\vec{g}_{\rm MI}(g_N)$ is one-to-one. Then since the Newtonian acceleration $g_N(r)$ goes to zero for both large and small radii,  the $g2$-space curves $\mathcal{C}$ of MOND modified inertia are closed curves with zero area. They are universally --- i.e. independent of the underlying baryonic matter distribution and independent of the details of the interpolation function --- neutral geometries according to the classification of table \ref{Table:geometries}. 
The MOND modified inertia curve in Fig.~\ref{Fig:geometries} (solid grey line) is computed from an infinitely thin exponential disk for the baryonic mass as indicated on the figure while that in Fig.~\ref{Fig:MONDcurves} is computed from a sum of 3 Miyamoto-Nagai disks as discussed in Sec.~\ref{sec:num}.

\bigskip
\subsection{The Deep Mondian Regimes at Large and Small Radii}
At large radii the solenoidal field $\vec{S}$ in Eq.~\eqref{MONDgrav} vanishes faster than the Newtonian acceleration $\vec{g}_N$ and so can be neglected. 
More precisely the unitless Newtonian acceleration $y=g_N/g_0$ falls off as $y \sim 1/r^2$ while the solenoidal field $\vec{S}$ vanishes faster than $1/r^3$ as shown in \cite{Bekenstein:1984tv}.  Since the inverse interpolation function in the deep mondian limit $y\to 0$ behaves as $\nu(y)\to y^{-1/2}$  the large radius limit of the Mondian acceleration is related to the Newtonian as
\begin{equation}
 g_{\rm M}\to \sqrt{g_N g_0}   , \quad {\rm for} \quad r\to \infty
\label{Eq: MONDliminf}
\end{equation}
The $g2$-space curves and rotation speed curves of MOND modified inertia and MOND modified gravity therefore always coincide at large radii. This is seen in the top left panels of Fig.~\ref{Fig:geometries} and  Fig.~\ref{Fig:MONDcurves} where 
all curves coincide at large radii, coresponding to the $g_{\rm N}\to 0$ parts of the $\mathcal{C}^+$ curve segments (the upper curve segments).

\bigskip
{\bf Analytical approximations:} 
To study the deep mondian limit  $y\to 0$ at small radii it is instructive to consider an analytic approximation to MOND modified gravity for  
infinitely thin disks. In particular we consider an exponential disk with surface mass density $\Sigma(r)=\Sigma_0 e^{-r/r_d}$. 
In this case an approximate expression for the resulting centripetal acceleration $g_{\rm BM,r}$ in MOND modified gravity was given by Brada and Milgrom in \cite{Brada:1994pk}. Taking Eq.~\eqref{Eq: MONDI} as an approximation for the MOND modified gravity acceleration outside the disk and then taking into account the discontinuity of the $z$-component of the acceleration inside the disk, one finds the radial acceleration in the plane of the infinitely thin disk: 
\begin{align}
 g_{\rm BM,r}(g_{\rm N}, r)= \nu(\frac{g_{\rm N}^+}{g_0})  g_{\rm N,r} , \quad g_{N^+}=\sqrt{g_{\rm N,r}^2+ (2 \pi G \Sigma(r))^2}  ; 
 \label{Eq:MONDgrav}
\end{align}
where  $g_{N^+}$ is the the total acceleration in the disk, including the discontinuity in the $z$-component by taking the limit $z\to 0^+$ from the upper half plane.
Given the limiting behaviour of the density $\Sigma(r)$ at large and small radii we find the centripetal acceleration in the two deep MOND regimes to be   
\begin{align}
 g_{\rm BM,r}&= \nu(\frac{g_{\rm N}^+}{g_0})  g_{\rm N,r} \to \sqrt{g_{N,r} g_0}   , \quad {\rm for} \quad r\to \infty
\\ \nonumber
g_{\rm BM,r}&=\nu(\frac{g_{\rm N}^+}{g_0})  g_{\rm N,r}  \to \sqrt{\frac{g_0}{\Sigma_0}} g_{N,r}, \quad {\rm for} \quad   r\to 0 
\label{Eq:testlim}
\end{align} 
At large radii this MOND modified gravity approximation coincides with MOND modified inertia as it should while at small radii it is reduced by a factor $\sqrt{\frac{g_{N,r}}{\Sigma_0}}<1$ relative to MOND modified inertia. 
In the deep MOND regime at small radii, the MOND and Newtonian accelerations are linearly related as opposed to the square root relation in the deep MOND regime at large radii. We show this MOND modified gravity approximation in Fig.~\ref{Fig:geometries} as the dotted acceleration (left panel) and speed curves (right panel) and in Fig.~\ref{Fig:MONDcurves} as the dotted black curve. The square root and linear behaviour of $g_{\rm BM,r}$ as a function of $g_{N,r}$ in the deep Mondian regimes at small and large radii respectively are clearly seen in the figures. It follows that in this approximation MOND modified gravity curves are cored with $r_{\rm tot}>r_N$, $\mathcal{C}^{+}>\mathcal{C}^{-}$ and  $\mathcal{A}(\mathcal{C})>0$.

It is also possible to start from a spherical potential for which the MOND modified gravity acceleration coincides with the MOND modified inertia exactly. By adding an axisymmetric perturbation one can compute the solenoidal field $\vec{S}$ as a function of this perturbation. This is done in \cite{Ciotti:2005bi}. We now discuss how the cored geometry of MOND modified gravity arises beyond the infinitely thin disk approximation or the approximation of a nearly spherical mass distribution.

\bigskip
{\bf Quasilinear MOND modified gravity: } 

Before solving the full MOND modified gravity geometries it is also instructive to consider the quasilinear version of MOND modified gravity (QUMOND) \cite{Milgrom:2009ee} to see the origin of the cored geometry.
The QUMOND acceleration $\vec{g}_{QM}$ is obtained by starting from the (pristine) MOND modified inertia acceleration $\vec{g}_{\rm MI}$ in the right hand side of Eq.~\eqref{Eq: MONDI} and then adding a solenoidal field $\vec{\sigma}=\nabla \times \vec{A}$ to enforce that the resulting QUMOND  acceleration $\vec{g}_{QM}$ has zero curl. this ensures that the QUMOND acceleration is derivable from a potential:
\begin{equation}
\vec{g}_{QM} =\nu(\frac{g_N}{g_0}) \vec{g}_{\rm N} +  \vec{\sigma}   , \quad \nabla \times \vec{\sigma} = - \frac{\nu'}{g_0}  \nabla g_N \times \vec{g}_{\rm N}  , 
\label{Eq:QMOND}
\end{equation}
where the curl of $\vec{\sigma} $ follows from the requirement $ \nabla \times \vec{g}_{QM}=\vec{0}$ such that $\vec{g}_{QM}=-\nabla  \psi_{QM}$. This allow us to determine $\nabla \times \vec{\sigma}$  straightforwardly in terms of the newtonian potential.
.

\begin{figure}[htp!]
	\centering
		\includegraphics[width=0.3\textwidth]{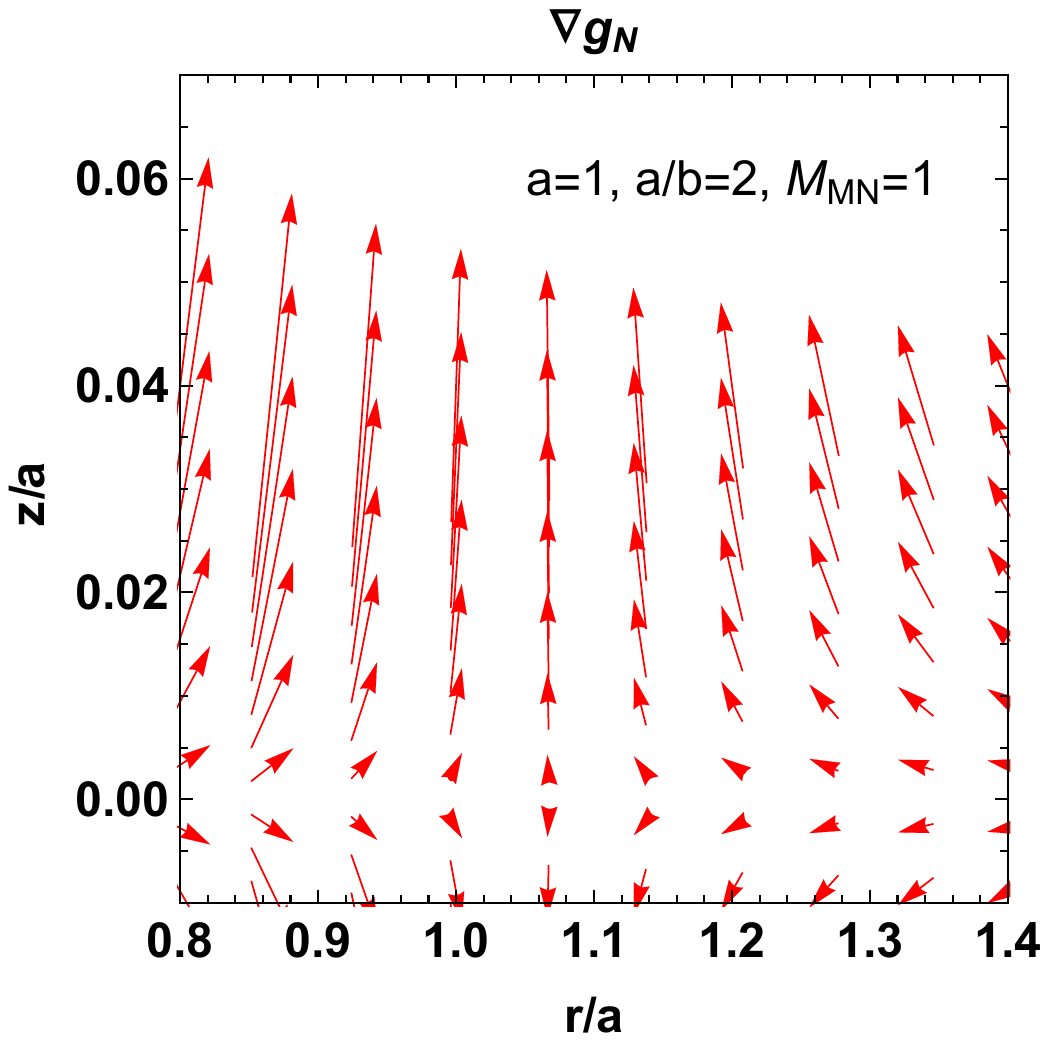}
		\includegraphics[width=0.3\textwidth]{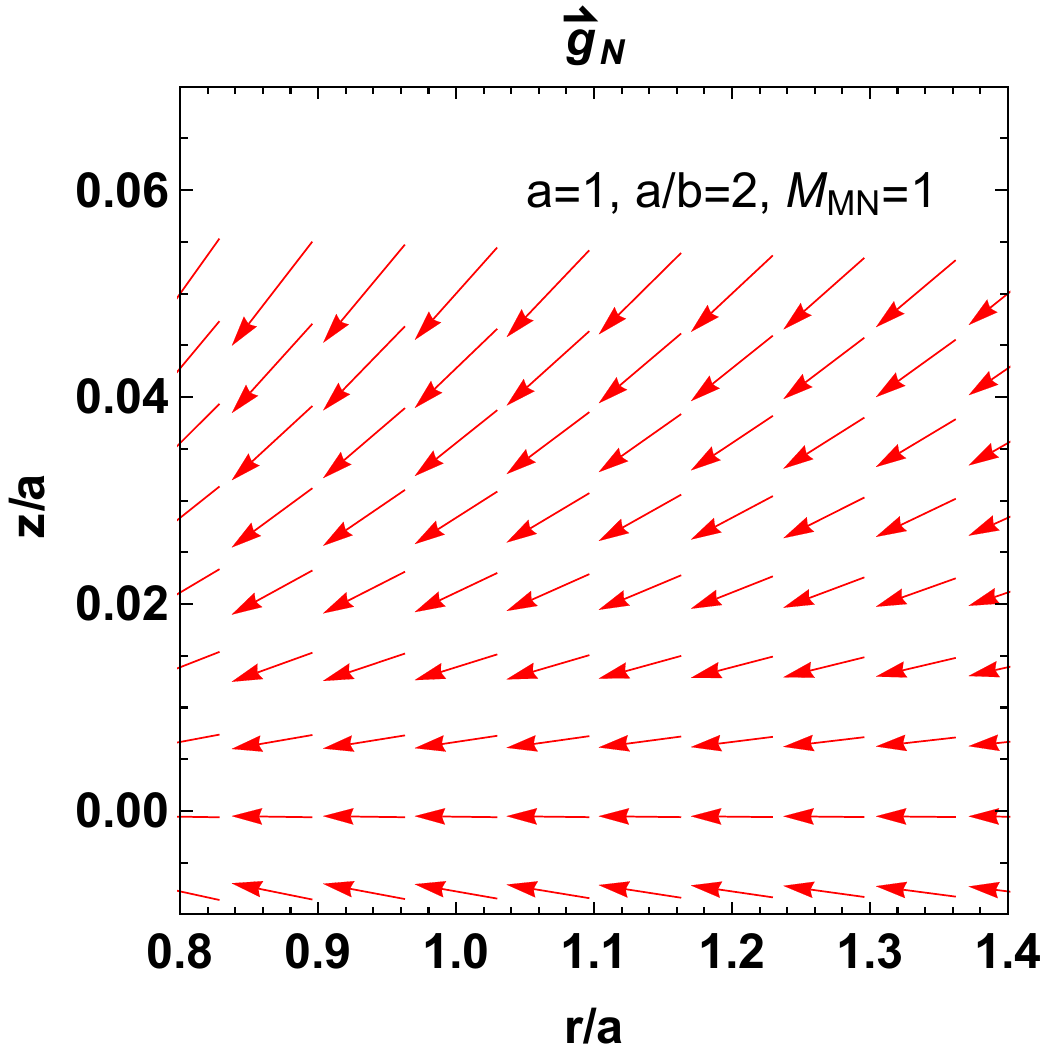}
		\includegraphics[width=0.37\textwidth]{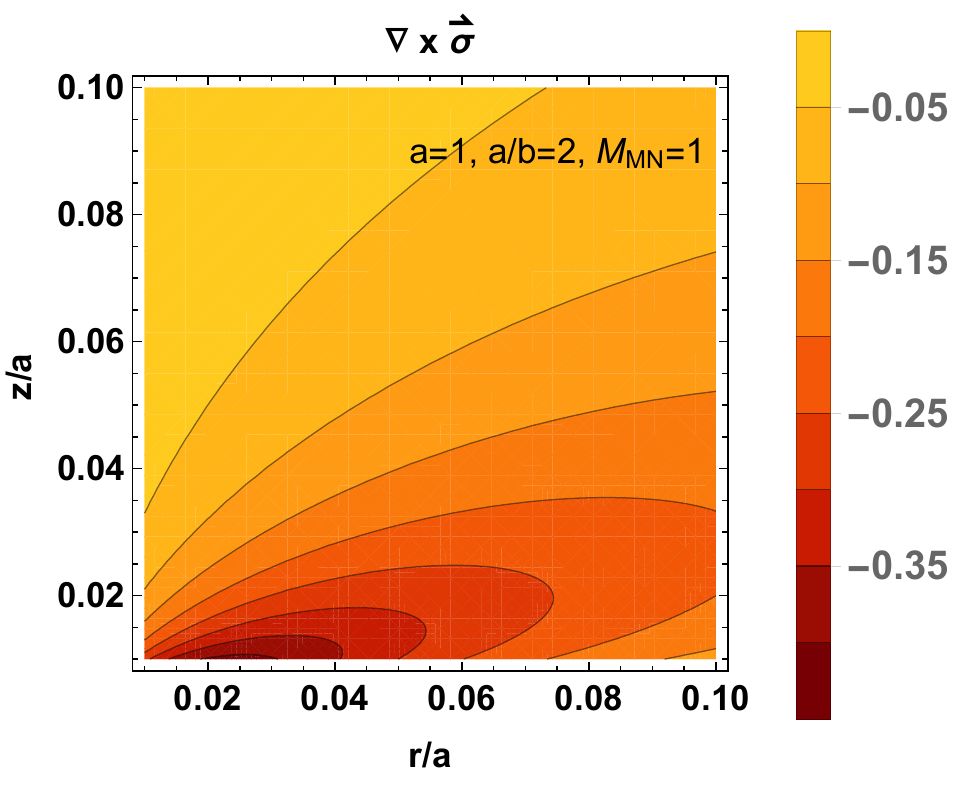}
		\caption
	{The gradient of the norm of the Newtonian acceleration  $\nabla g_N$ (left panel), the Newtonian acceleration vector $\vec{g}_{\rm N}$ (middle panel) and the curl of the solenoidal field $\nabla \times \vec{\sigma}$ (right panel) in QUMOND for a Miyamoto-Nagai disk model with mass parameter $M_{\rm MN}=1$, scale height $b=0.5$ and scale length $a=1$. From the sign of $\nabla \times \vec{\sigma}$ we infer that the centripetal component of the curl acceleration is in the opposite direction of the Newtonian acceleration vector $\vec{g}_{\rm N}$ in the disk near $z=0$.}
	\label{Fig:MONDcurls}
\end{figure}

Due to the axisymmetry of the matter distribution, the curl $\nabla \times \vec{\sigma}$ is purely in the azimuthal direction and 
using the second identity in Eq.~\eqref{Eq:QMOND} we can also determine the sign of  $\nabla \times \vec{\sigma}$ in the azimuthal direction to be negative. This is shown in Fig.~\ref{Fig:MONDcurls} for a Miyamoto-Nagai (MN) potential given in Eq.~\ref{Eq:MNpot} and reviewed in Sec.~\ref{sec:num}. The scale length $a$, scale height $b$ and mass scale $M$ of the MN model used in the figure are $a=1, b=0.5$ and $M=1$ (bottom left). The negative sign arises as follows:  
For $z\geq 0$ the gradient $\nabla g_N$ is dominantly in the $z$-direction for a non-spherical, axisymmetric mass distribution (It is zero in the plane by symmetry unless the disk is infinitely thin) while $\vec{g}_N$ is dominantly in the negative $r$-
direction. This is illustrated for the MN model in Fig.~\ref{Fig:MONDcurls} in the left and middle panels respectively. It is particularly clear at the line of maximum radial acceleration $\partial_r g_N=0$ near $r=1.1$ where $\nabla g_N$ points entirely in the positive $z$-direction while $\vec{g}_N$ is dominantly in the negative radial direction. 
Then since $\nu' <0$ in the mondian regimes at small and large radii we have that  $\nabla \times \vec{\sigma}$ is in the negative azimuthal direction as seen in the right hand panel of the figure and consequently $\vec{\sigma}$ circulates in the counterclockwise direction in the $(r,z)$-plane. 

At large radii $\vec{\sigma}$ is negligible but at small radii the radial component of $\vec{g}_{QM}$ is reduced as compared to  the radial component of $\vec{g}_{MI}$ because the radial acceleration from the solenoidal field $\sigma$ is in the opposite direction to the Newtonian acceleration $\vec{g}_N$.
It follows that the QUMOND modified gravity geometries are cored, using the classification of table \ref{Table:geometries}.

\bigskip
{\bf Solenoidal field in MOND modified gravity: } 
The arguments above for QUMOND may also be applied to the solenoidal field $\vec{S}$  of the Bekenstein-Milgrom MOND modified gravity (note the opposite sign convention!). The curl $\nabla \times \vec{S}$ may be expressed from either of the two identities in Eq.~\ref{MONDgrav}, using $\nabla \times   \vec{g}_{\rm N} =\nabla \times   \vec{g}_{\rm M}=0 $  as
\begin{equation}
 \nabla \times \vec{S} = \mu'  \nabla g_M \times \vec{g}_M =  -\frac{\nu'}{ \nu g_0}  \nabla q \times \vec{q} 
\label{MONDgrav>}
\end{equation}
Due to the axisymmetry, the curl $\nabla \times \vec{S}$ is again purely in the azimuthal direction
We may argue that the sign of $\nabla \times \vec{S}$ is in the negative azimuthal direction such that $\vec{S}$ circulates in the counterclockwise direction by repeating the arguments for QUMOND:
At large radii the solenoidal field is negligible 
and as we move towards smaller radii we approximate  $\vec{q}\simeq \vec{g}_N$ in the above formula and repeat the arguments above for QUMOND, i.e.  near the $z=0$ plane the radial acceleration is reduced as compared to that in MOND modified inertia. 
It follows that the MOND modified gravity geometries are cored, like in QUMOND, using the classification of table \ref{Table:geometries}. 
Example numerical solutions of Eq.~\eqref{Eq:MONDgravstart}, using the {\tt N-MODY} code, are shown in the top left panel of Fig.~\ref{Fig:MONDcurves} (coloured dotted curves)  
as a function of the parameter $b$ controlling the departure from the spherical limit. We now discuss these numerical solutions.

\subsection{Numerical Solutions of MOND modified gravity}
\label{sec:num}
In order to study numerical solutions of the full MOND modified gravity geometries in $g2$-space, it is useful to consider Miyamoto-Nagai potentials which interpolate between axisymmetric potentials and spherical potentials. The Miyamoto-Nagai potential $\Phi_{MN}$ is given by
\begin{align}
\Phi_{MN}(r,z)=\frac{-G M_{MN}}{\sqrt{r^2+(a+\sqrt{z^2+b^2})^2}}
\label{Eq:MNpot}
\end{align} 
	where $M_{MN}$ is a mass parameter, $b$ is the vertical scale height, and $a$ is the radial scale length \cite{Miyamoto:1975zz}. 
In the limit $a=0$ the potential is spherically symmetric and reduces to the Plummer potential. For $a\neq 0$ the potential is axisymmetric (and symmetric about the $z=0$ plane) and in the limit $b=0$ the potential reduces to the potential of an infinitely thin Kuzmin disk for which the MOND modified gravity solution is known exactly.
 
However, instead of a single MN model we use a sum of 3 Miyamoto-Nagai potentials (3MN models) which can be used to approximate the potential of both thick and thin exponential disks \cite{3MN}, with densities of the form 
		\begin{gather}
			\rho(r,z)=\rho_0 \exp\left(-r/r_d\right) \exp\left( -|z|/z_0\right),  \rho_0=\frac{M}{4\pi h_z r_d^2}
		\end{gather}
where $r_d $ the radial scale length, $z_0$ the vertical scalelength. 
This allows us to compare numerical solutions of MOND modified gravity to both the Brada-Milgrom approximation for infinitely thin exponential disks and MOND modified inertia in the spherical limit. The 3MN potentials we use are therefore of the form:
\begin{align}
\Phi_{3MN}(r,z)=\sum_{i=1}^3 \frac{-G M_{MN,i}}{\sqrt{r^2+(a_i+\sqrt{z^2+b_i^2})^2}}
\label{Eq:MNpot}
\end{align} 

We discuss details of the 3MN approximation in the appendix.

\begin{figure}[htp!]
	\centering
\includegraphics[width=0.38\textwidth]{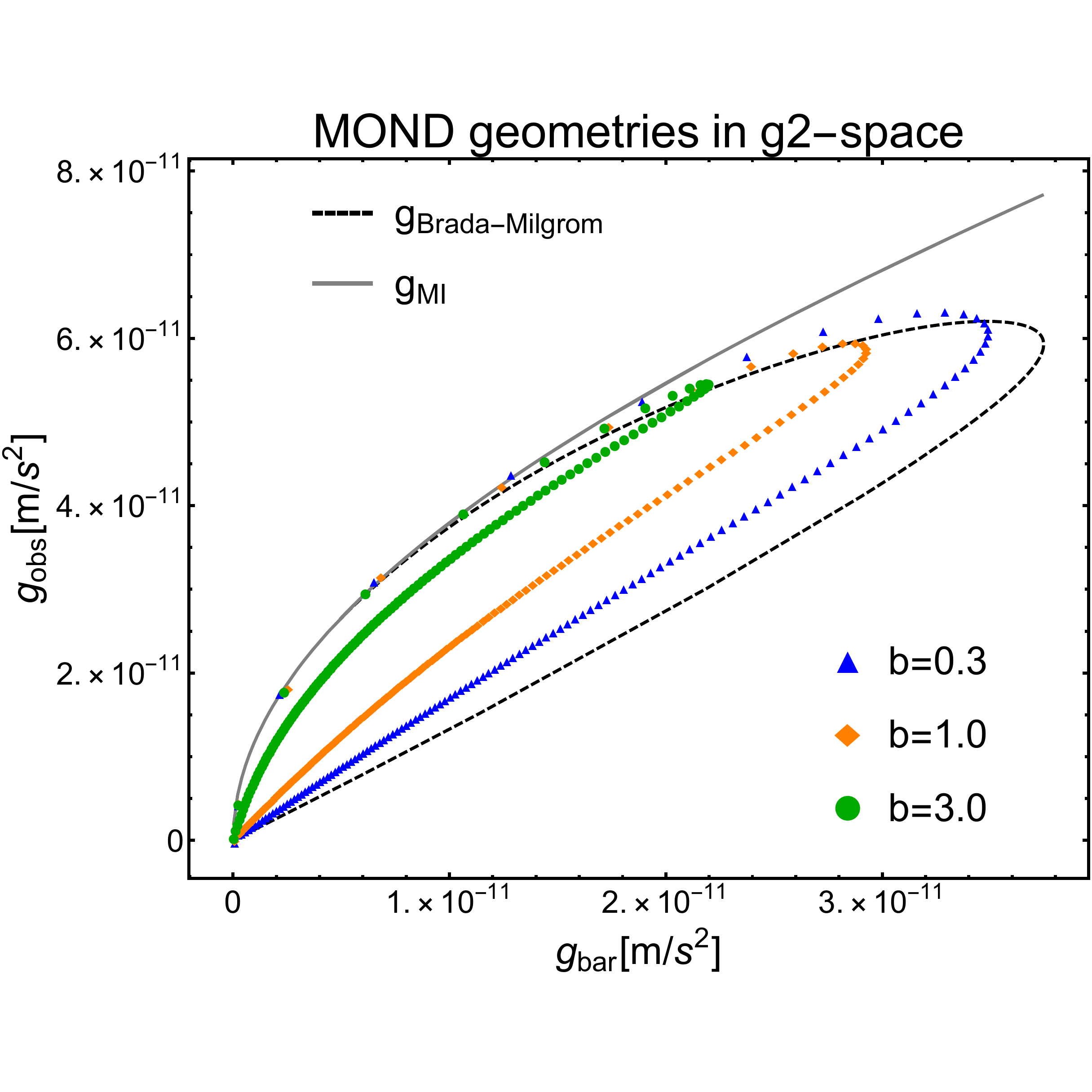}
\includegraphics[width=0.38\textwidth]{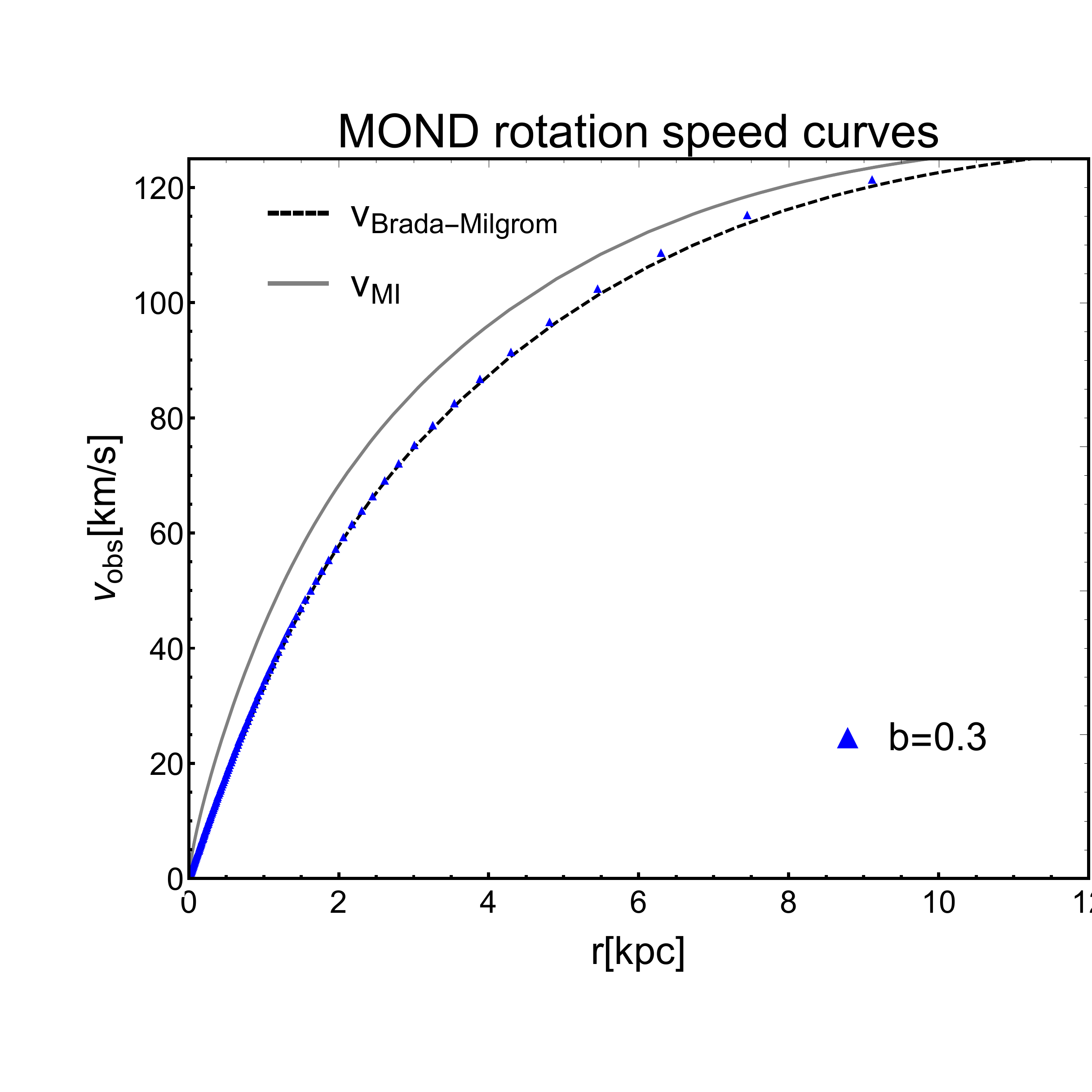}
	\includegraphics[width=0.4\textwidth]{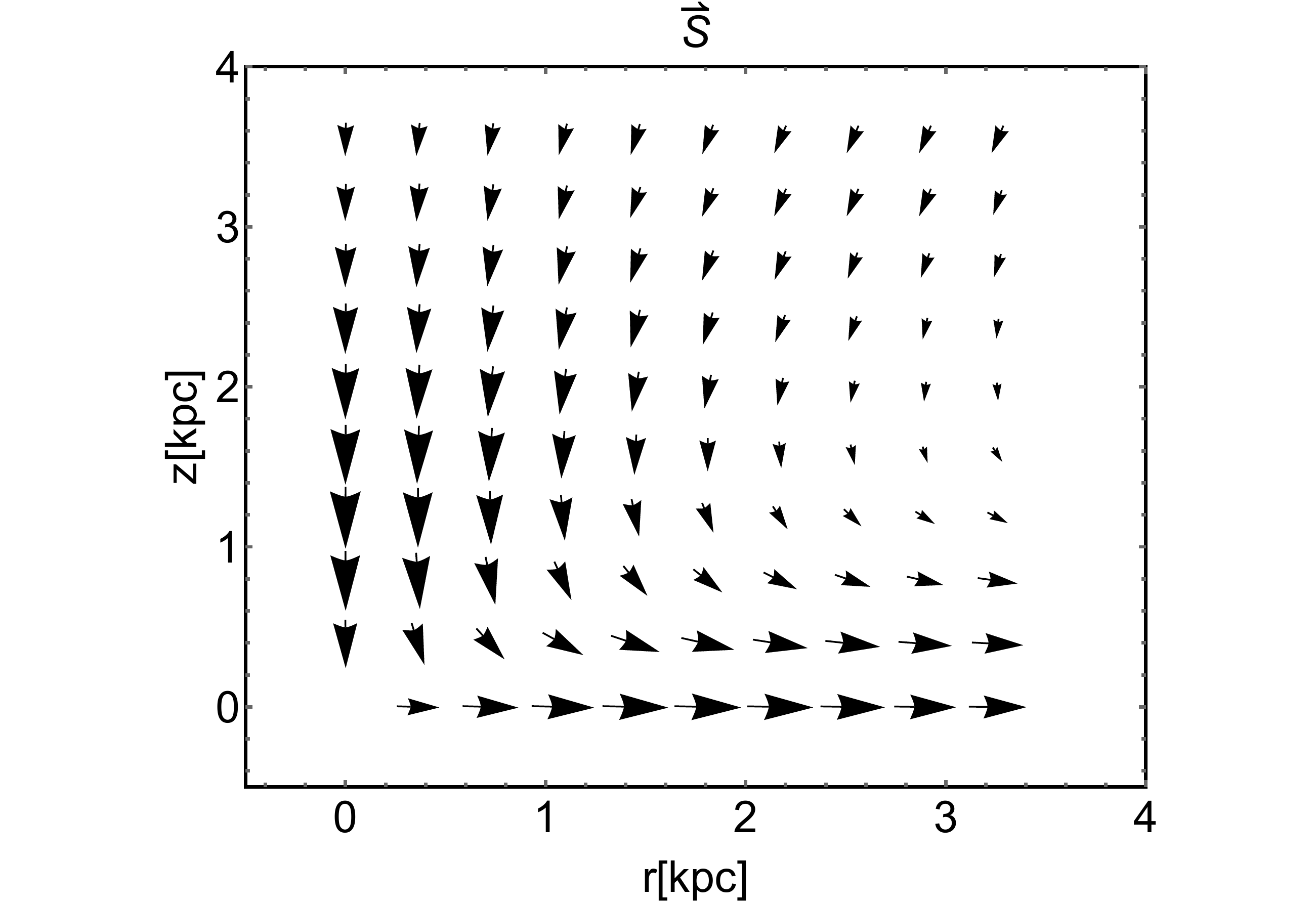}
		\includegraphics[width=0.4\textwidth]{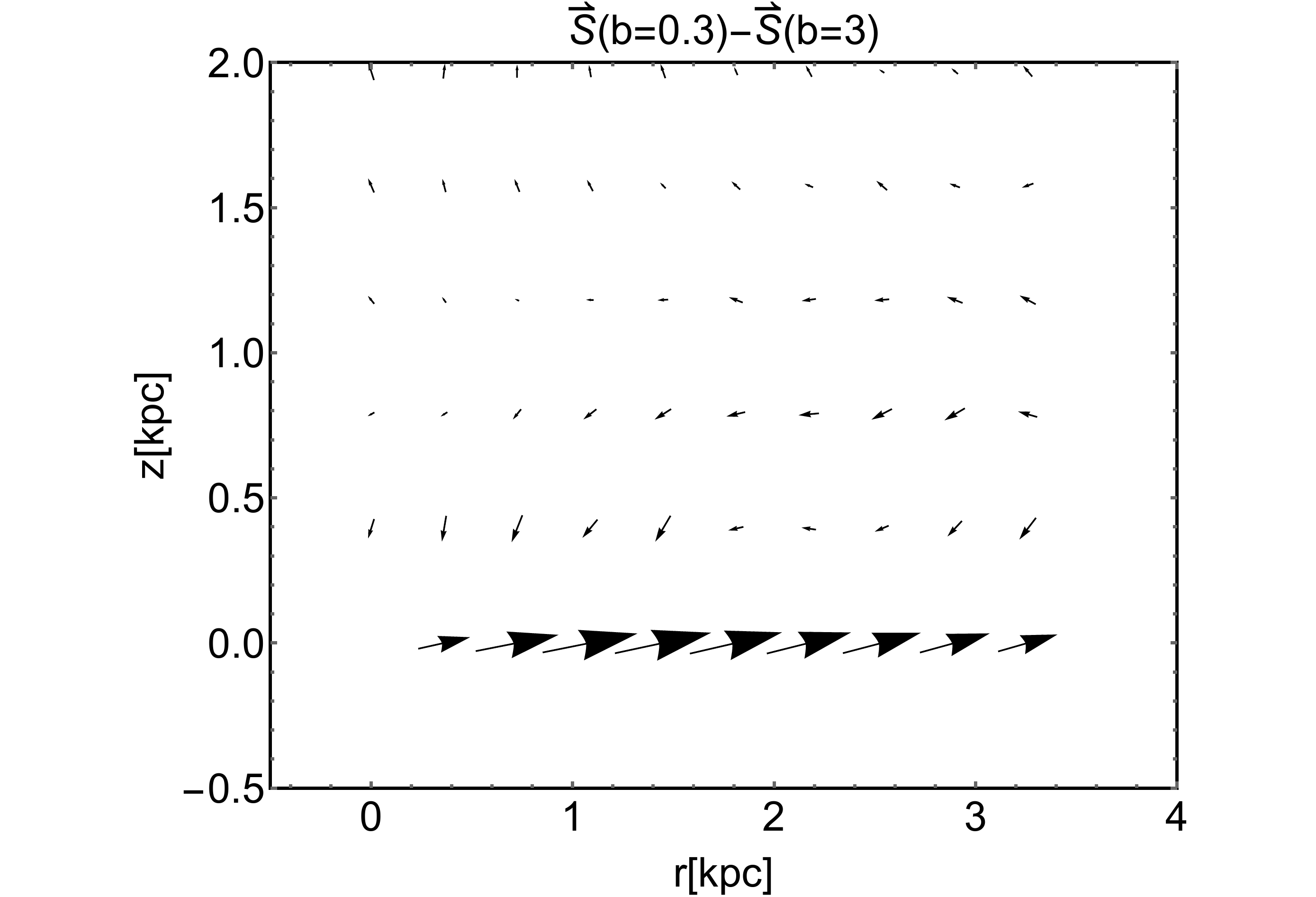}
	\caption
	{ {\it Top left:} Centripetal acceleration curves in $g2$-space of MOND modified inertia (solid grey curve) and MOND modified gravity in the Brada-Milgrom approximation (black dotted curve) with acceleration scale $g_0=1.2\times 10^{-10} \frac{m}{s^2}$ for a 3MN model of an infinitely thin exponential disk galaxy with $M=1.2 \times 10^{10} M_{\odot}$ and a scalelength of $r_d=3.5$ kpc.The 3MN parameters are given in the appendix. 
	 The dotted colored curves are the full MOND modified gravity solution of the same model but varying the scale height parameter of the 3MN model $b=3$ (green), $b=1$ (red) to $b=0.3$ (blue) ( $b\to 0$ is the disk limit  and $b\to \infty$ is the spherical limit)
	 {\it Top right panel:} The corresponding rotation curves showing the full MOND modified gravity solution only for $b=0.3$. 
		{\it Bottom left Panel}: The corresponding curl field $\vec{S}$ for the $b=3$ case showing how the curl field at $z=0$ is opposite in direction to the Newtonian one and reduces the radial MOND modified gravity acceleration compared to the MOND modified inertia. 
		{\it Bottom right Panel}: The difference in curl fields for the most disk like (b=0.3) and most spherical (b=3) parameter values, showing that the solenoidal field reduces the radial MOND modified gravity acceleration in the  $z=0$ plane most for the most disk-like mass distributions.  
	}
	\label{Fig:MONDcurves}
\end{figure}

\bigskip
As an example we start from an infinitely thin exponential disk galaxy with mass of $M=1.2 \times 10^{10} M_{\odot}$ and a scalelength of $r_d=3.5$ kpc and  take $g_0=1.2\times 10^{-10} \frac{m}{s^2}$. The corresponding 3MN model parameters $M_{3MN}, b=0, a_i, M_i$ are given in the appendix and used to compute the MOND modified inertia $g_{MI}$ curve (solid grey line) and MOND modified gravity in the Brada Milgrom approximation $g_{BM}$ (dotted black line) in the top left panel of Fig.~\ref{Fig:MONDcurves}. By dialling the parameter $b$ up we can make the mass distribution more spherical with $a/b\to 0$ the spherical limit  (corresponding do increasing the scale height $z_0$ in the thick exponential disk above) . 
The colored dotted curves show the full solution of the MOND modified gravity curve, using the static solver of the {\tt N-MODY} code for the 3MN models with increasing values of $b$ from $b=0.3$ (blue), $b=1$ (orange) to  $b=3$ (green). 

The figure demonstrates the properties discussed previously:  The MOND modified inertia and MOND modified gravity curves coincide at large radii but the MOND modified gravity curves are more cored at small radii and the core grows in proportion to the departure from spherical symmetry (as measured by the ratio $a/b$ here) towards the Brada-Milgrom approximation in the thin disk limit $a/b\to \infty$. We show the rotation speed curves corresponding the MOND modified inertia and Brada-Milgrom curves along with the full numerical solution for $b=3$ it the top right panel. 
Finally it is seen how the approach to zero of the curves follow the limiting behaviour given in Eq.~\eqref{Eq: MONDliminf} and Eq.~\eqref{Eq: MONDliminf}. 
We also note that if we increase the mass of the disk, such that the accelerations reach the Newtonian regime, simply means the curves are stretched and the core effect at small radii is less pronounced.

In the bottom left panel we show the corresponding solenoidal field $\vec{S}$ for the $b=3$ example. The curl field indeed circulates in the counterclockwise direction. Finally in the bottom right hand panel of the figure we show the difference between the $b=0.3$ and $b=3$ curl fields is positive near $z=0$ and thus the least spherical potential leads to the biggest curl field along the $z=0$ plane and thus the biggest core effect.  
We can therefore take the Brada-Milgrom approximation as a good approximation of MOND modified gravity for thin disks. For more spherical mass distributions the Brada-Milgrom will overestimate the core so for a given galaxy we may take the Brada-Milgrom formula and apply it to a thin disk of equivalent mass to get an upper limit on the core produced by MOND modified gravity in the galaxy.

In summary we have shown that MOND modified gravity leads to cored $g2$-space geometries (and corresponding rotation curves) for isolated galaxies and that the core effect is then smallest for the most spherical systems like e.g. dwarf spheroidals. MOND modified inertia, which is also the limit of MOND modified gravity for fully spherical baryonic mass distributions, lead to neutral geometries.

\section{SPARC galaxy data in $g2$-space}
We now compare MOND models with cuspy and cored rotation curve data in the {\tt SPARC} database \cite{Lelli:2016zqa}.
Since the geometry of MOND modified gravity is cored relative to MOND modified inertia it yields worse fits to cuspy galaxies. We can therefore use the fits of MOND modified inertia to cuspy galaxies to provide a simple upper limit on the goodness of fit for MOND modified gravity to cuspy galaxies. 
Conversely from Fig.~\ref{Fig:MONDcurves} we can use the Brada-Milgrom approximation as an upper limit on the maximal cores of MOND modified gravity. Below we show examples of galaxies where the cores of MOND modified gravity in the Brada-Milgrom approximation are insufficient.  
This analysis serves to demonstrate the quantitative relevance of the cusp-core challenge for MOND. 

The summary of our data analysis  below follows that in \cite{Frandsen:2018ftj}. The  {\tt SPARC} database provides the observed rotational velocities, $v_{\rm obs}(r_{j,G})$, as well as the inferred contributions to the baryonic velocity from the disk, bulge and gas in the galaxy, $v_{\rm disk}(r_{j,G})$, $v_{\rm bul}(r_{j,G})$, $v_{\rm gas}(r_{j,G})$ with $r_{j,G}$ the $j$'th measured radius in the galaxy $G$. The accelerations $g_{\rm obs}(r_{j,G})$ and $g_{\rm bar}(r_{j,G}),$ are then
\begin{equation}
\begin{split}
&g_{\rm obs}(r_{j,G}) = \dfrac{v_{\rm obs}^{2}(r_{j,G})}{r_{j,G}},\\
&g_{\rm bar}(r_{j,G}) = \frac{|v_{\rm gas}(r_{j,G})| v_{\rm gas}(r_{j,G}) + \Upsilon_{\rm disk,G} |v_{\rm disk} (r_{j,G})v_{\rm disk} (r_{j,G})| + \Upsilon_{\rm bul,G}|v_{\rm bul}(r_{j,G})| v_{\rm bul}(r_{j,G})}{r_{j,G}},
\label{Eq:accerrors}
\end{split}
\end{equation}
where $\Upsilon_{\rm disk,G}$ and $\Upsilon_{\rm bulge,G}$ are unitless mass to light ratios. Following \cite{McGaugh:2016leg,Lelli:2017vgz,Lelli:2016zqa,Li:2018tdo} we will take the central values $\Upsilon_{\rm disk,G}= 0.5$ and $\Upsilon_{\rm bulge,G}= 0.7$ including a $25 \%$ relative uncertainty on both $\delta\Upsilon= 0.25 \Upsilon$. The SPARC data base also provides the corresponding (random) uncertainties $\delta {v_{\rm obs}}(r_{j,G})$, as well as the uncertainties $\delta i_G $ and $\delta D_G$ on the galaxy inclination angle $i_G$ and distance $D_G$. Following \cite{Lelli:2017vgz}~we further adopt a $10\%$ uncertainty on the HI flux calibration translating into a  $10\%$ uncertainty on the gas accelerations, i.e. $\delta g_{\rm gas}(r_{j,G})=0.1 g_{\rm gas}(r_{j,G})$. From these uncertainty contributions we can compute the full $  \delta {g_{\rm bar}},  \delta {g_{\rm obs}} $ uncertainties
\small
\begin{equation}
\begin{split}
\delta {g_{\rm obs}}(r_{j,G})  &= g_{\rm obs}(r_{j,G})  
\sqrt{\bigg[\dfrac{2\delta{v_{\rm obs}} (r_{j,G}) }{v_{\rm obs}(r_{j,G}) }\bigg]^{2} 
	+ \bigg[\dfrac{2\delta i_G}{\tan(i_G)}\bigg]^{2} 
	+ \bigg[\dfrac{\delta D_G}{D_G}\bigg]^{2}},
\\
\delta {g_{\rm bar}}(r_{j,G})  &= g_{bar}(r_{j,G})\sqrt{\bigg(\frac{\delta g_{\rm gas}(r_{j,G})}{g_{\rm bar}(r_{j,G})}\bigg)^2 + \sum_{k=disk,bulge}\bigg(\frac{v_{k}^2(r_{j,G}) \delta \Upsilon_{k,G}}{v_{bar}^2(r_{j,G})}\bigg)^2 },
\end{split}
\label{Eq:uncertaintiest}
\end{equation}
\normalsize
where we note that the inferred $g_{bar}(r_{j,G})$ are independent of distance $D_G$ and inclination angle $i_G$ \cite{Li:2018tdo}. We treat the uncertainties $\delta{v_{\rm obs}}$ as random Gaussian distributed uncertainties for each data point while the remaining uncertainties, $\delta i_G $, $\delta D_G$, $\delta \Upsilon_{\rm disk,bulge,G},\delta g_{gas}(r_{j,G})$ are systematic within each galaxy meaning they rescale all data points within a given galaxy in one direction.

The normalized residuals which enter into the $\chi^2$-function, taking into account the errors in both the observed and baryonic accelerations can be approximated as \cite{Orear:1981qt}
\begin{equation}
R(r_{j,G}) =\frac{\left(g_{\rm obs}  (r_{j,G})-g_{MI} (r_{j,G}) \right) }{ \sqrt{\delta {g_{\rm obs}}(r_{j,G})^2+ g'_{MI} (r_{j,G})^2  g_{\rm abs}(r_{j,G})^2}}
\label{Eq:residuals}
\end{equation}
where $ g'_{MI} (r_{j,G})= \frac{\partial {g_{MI} (g_{\rm bar}(r_{j,G}))}}{\partial g_{\rm bar}(r_{j,G})} $ is the derivative with respect to the baryonic acceleration.

\subsection{Data comparison in $g2$-space}

\begin{figure}[htp!]
	\centering
	\includegraphics[width=0.35\textwidth]{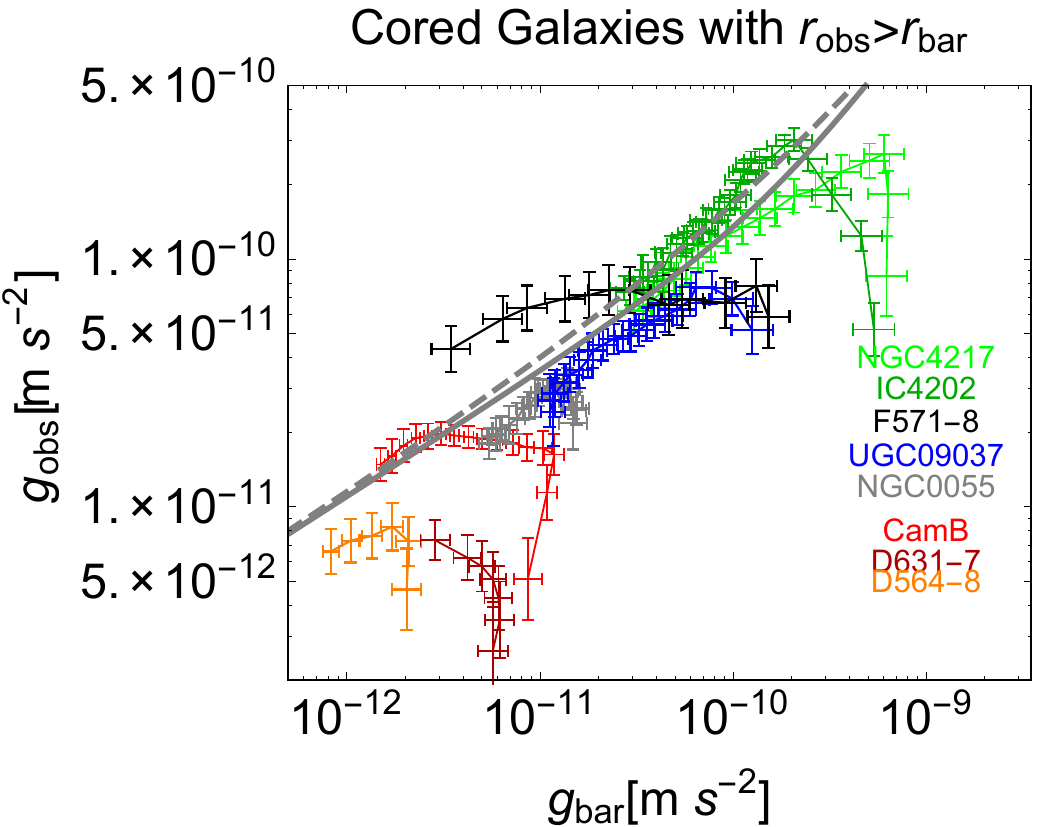}	
	\includegraphics[width=0.35\textwidth]{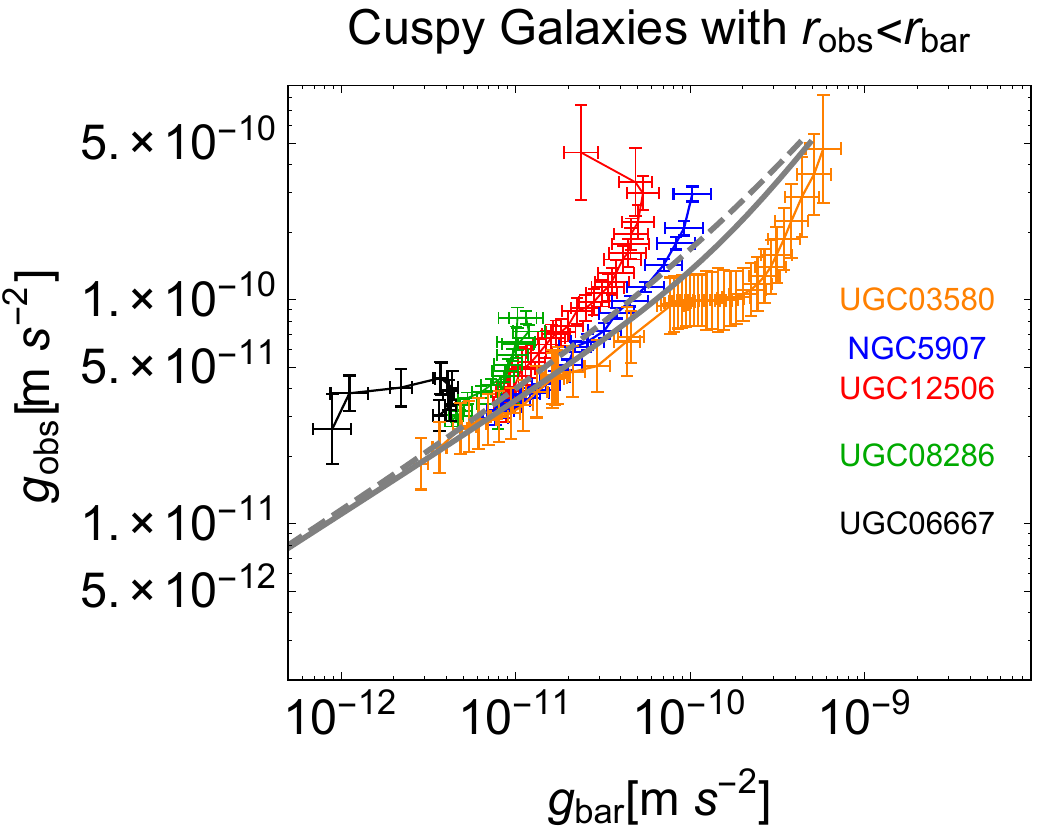}
	\includegraphics[width=0.35\textwidth]{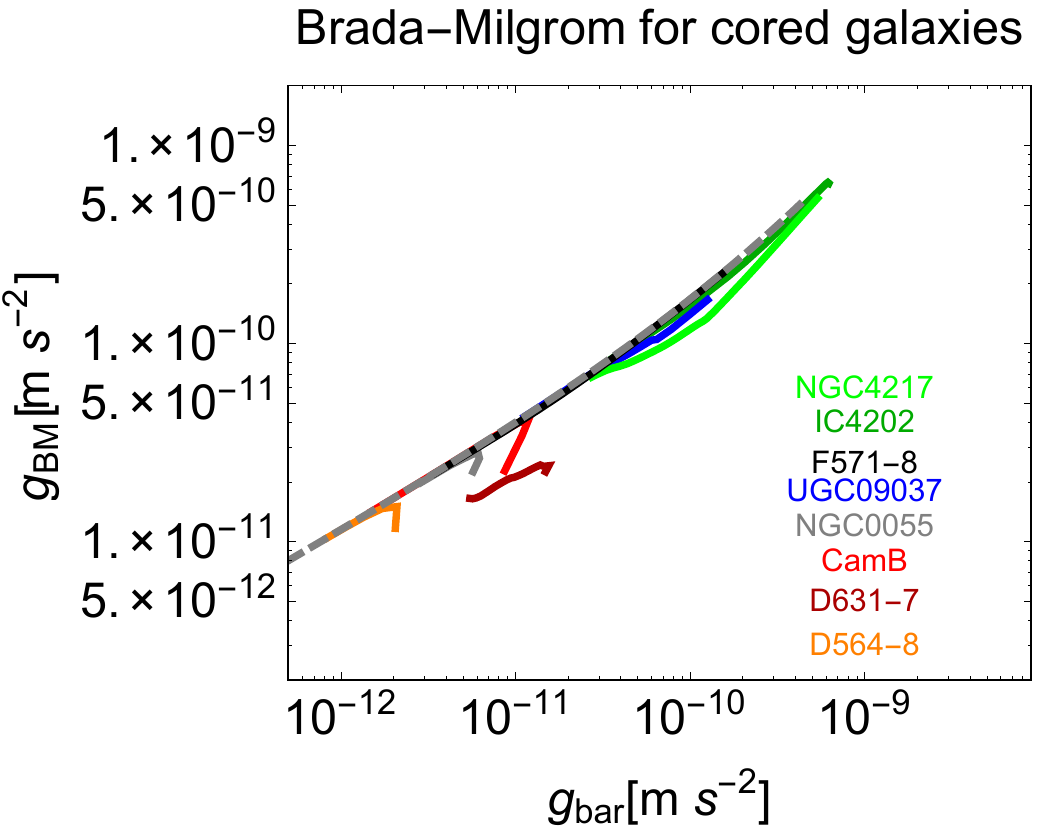}	
	\includegraphics[width=0.35\textwidth]{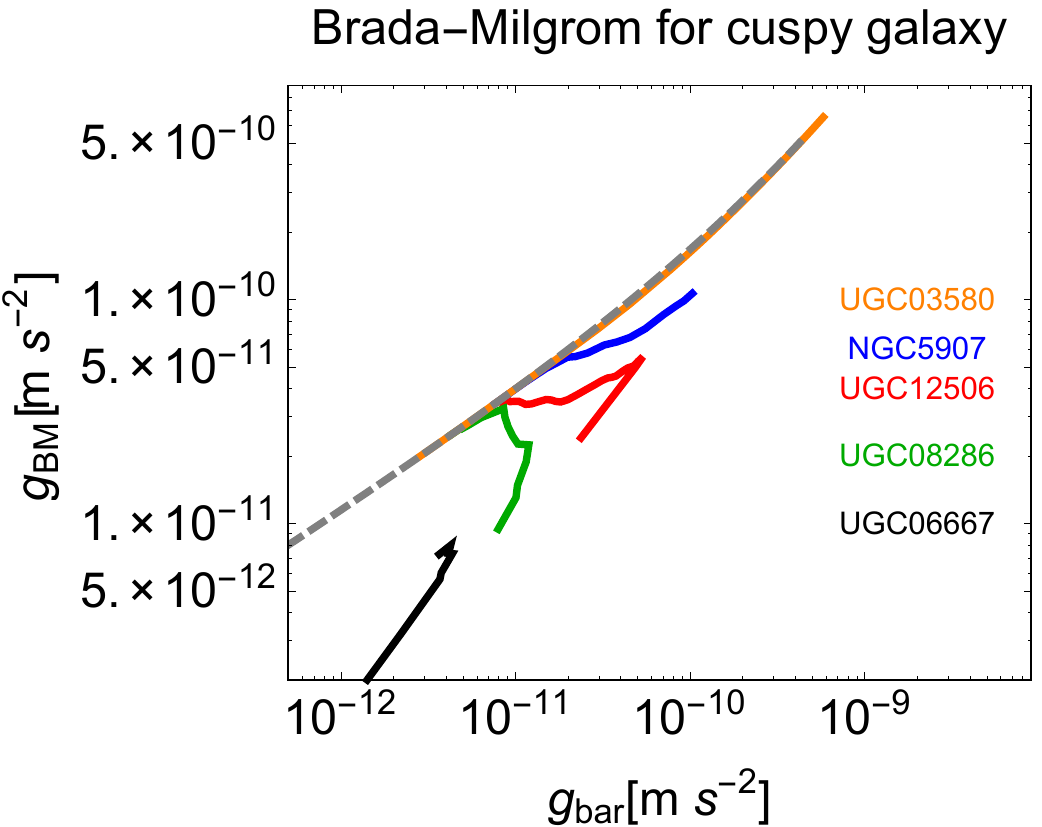}
	\includegraphics[width=0.35\textwidth]{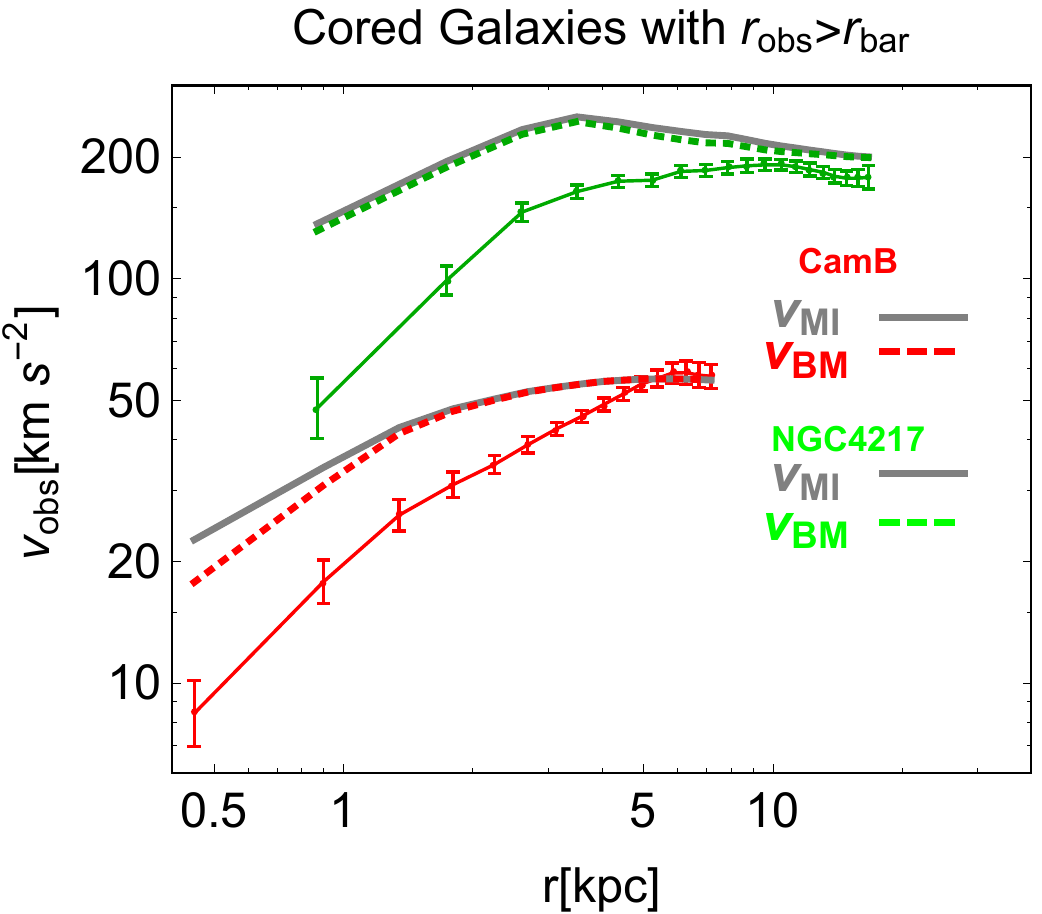}	
	\includegraphics[width=0.35\textwidth]{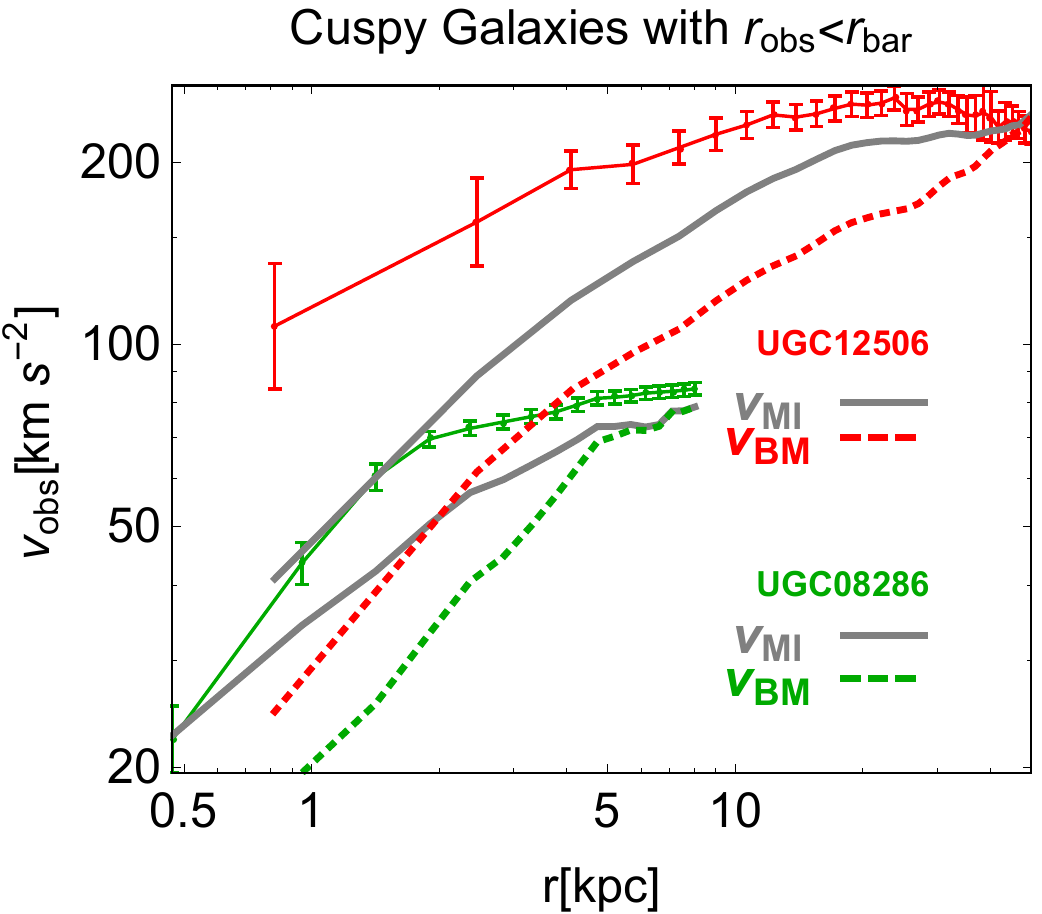}
	\caption
	{Top left: Acceleration curves of galaxies from SPARC in $g2$-space with cored geometry $r_{\rm obs}>r_{\rm bar}$. Top right: The same for cuspy galaxies with $r_{\rm obs}<r_{\rm bar}$. Also shown on both panels are the MOND modified inertia curves for the two considered interpolation functions in Eq.~\eqref{Eq:interpolationfunctions}. 
	Middle left: The corresponding model curves for the cored galaxies from MOND Modifed gravity curves using Brada-Milgrom approximation with the $g_{\rm bar}$ and surface density $\Sigma(r)$ values from SPARC. Middle right: The same as left but for the cuspy galaxies. For reference also the MOND modified inertia curve is again shown.  
	Bottom left: The corresponding rotation speed curve data for two of the cored galaxies, compared to the MOND modified inertia and MOND modified gravity curves. Bottom right: The same as left for two cuspy galaxies}
	\label{Fig: Galaxiescorecusp}
\end{figure}

On the top left panel of Fig.~\ref{Fig: Galaxiescorecusp} we show galaxies from {\tt SPARC} with cored $g2$-space geometries (color legend on figure identifies the individual galaxies) along with the modified inertia model with the interpolation functions in Eq.~\eqref{Eq:interpolationfunctions}  and with the best fit value $g_0=1.2\cdot 10^{-10} m/s^2$ from \cite{Lelli:2016zqa} (gray and gray dashed lines). 
The galaxies were selected as cored only by requiring  $r_{\rm obs,G}>r_{\rm bar,G}$ and further by having a large $\chi^2_G$ value for the fit of data from each galaxy to the MOND modified inertia curve \eqref{Eq:interpolationfunctions}. Below we discuss the fits in more detail.   
On the top right panel we show the same for galaxies with cuspy $g2$-space geometries selected as cuspy only by requiring  $r_{\rm obs,G}<r_{\rm bar,G}$, and again by further having a large $\chi^2_G$ for the fit of the data from each galaxy to the curve in Eq.\eqref{Eq:interpolationfunctions}.

On the middle row of panels we then show the predicted MOND modified gravity curves of the same cored and cuspy galaxies along with the MOND modified inertia curves for reference. We obtain the MOND modified gravity curves using the Brada-Milgrom approximation in Eq~\eqref{Eq:MONDgrav} with $g_N$ replaced by the data $g_{\rm bar}(r_{j,G})$ and with $\Sigma(r)$ replaced by the disk surface densities $\Sigma(r_{j,G})$ from {\tt SPARC}. 

It is visually clear that the MOND modified gravity curves yield poor fits to both the cored and cuspy galaxies as we quantify further below.  
For the cuspy galaxies the MOND modified gravity is by nature of its cored geometry a worse description than MOND modified inertia.  For the cored galaxies the MOND modified gravity curves are everywhere too close to the MOND modified inertia limit to describe the cored geometry of the data.
In particular the MOND modified gravity curves are more cored for the cuspy galaxies than they are for the cored ones,  because the latter have larger stellar surface mass densities. 

On the bottom row of panels we show the corresponding rotation speed curves for two of the cored and two of the cuspy galaxies compared to the predictions from MOND modified inertia and the Brada-Milgrom approximation of MOND modified gravity. 

\begin{figure}[htp!]
	\centering
        \includegraphics[width=0.35\textwidth]{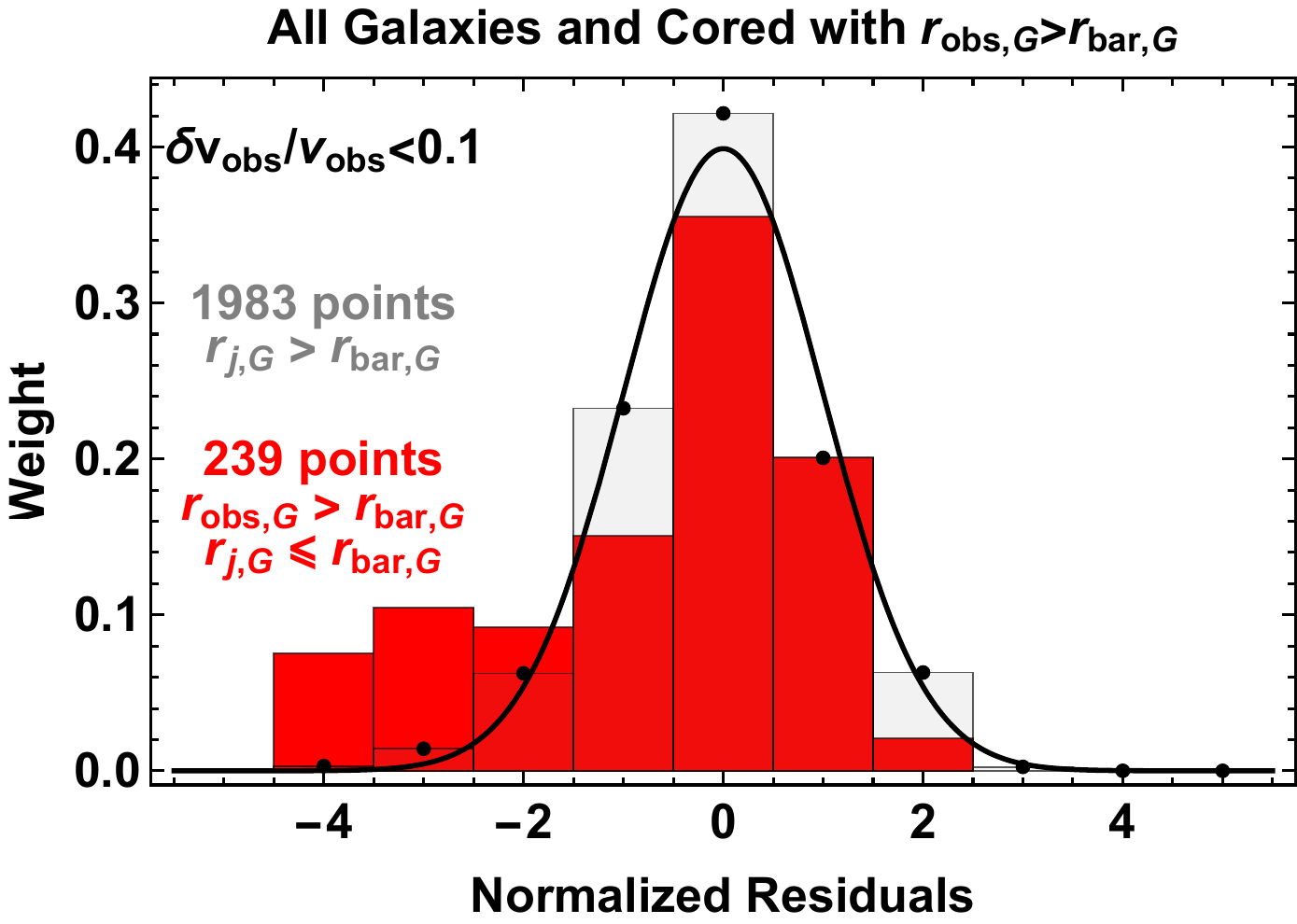}	
	\includegraphics[width=0.35\textwidth]{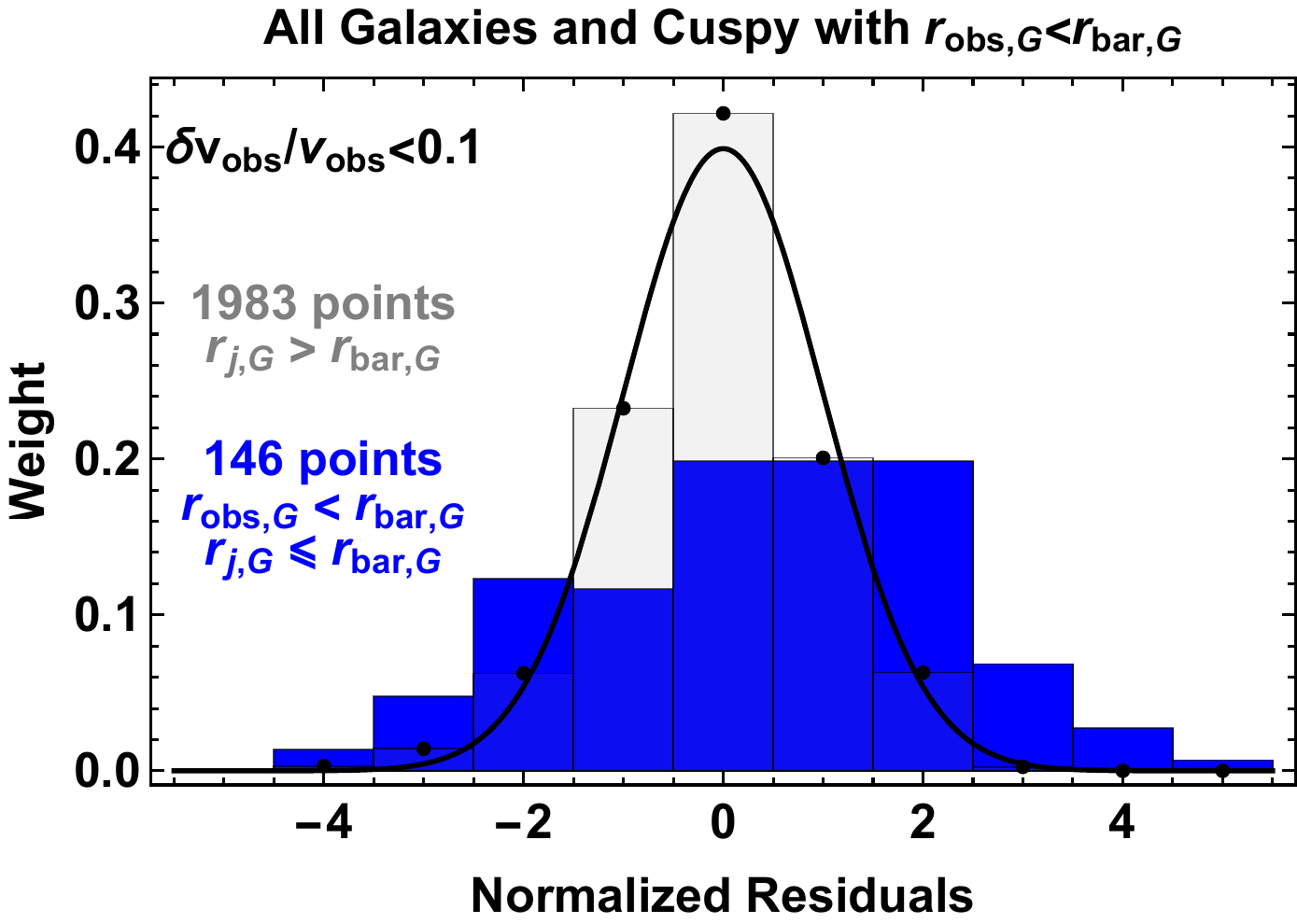}
	\caption
	{Left: Distribution of the normalized residuals $R(r_{j,G})$ in Eq.~\eqref{Eq:residuals} of SPARC data with respect to MOND modified inertia with $g_0=1.2\cdot 10^{-10} m/s^2$. The gray histograms with black dots on each bin top shown on both panels are the 1983 SPARC data points at large radii, i.e. $r_{\rm j,G}>r_{\rm bar,G}$ after imposing the basic quality criteria and $\delta v_{\rm obs}/v_{\rm obs}<0.1$ cut employed in \cite{McGaugh:2016leg,Lelli:2017vgz,Lelli:2016zqa}. Also shown on the left panel are normalized residuals of  data points at small radii $r_{\rm j,G}\leq r_{\rm bar,G}$ from the (cored) galaxies only with $r_{\rm obs,G}>r_{\rm bar,G}$ (red histogram). Right: On the right panel we also show  points at small radii $r_{\rm j,G}<r_{\rm bar,G}$ from (cuspy) galaxies only with  $r_{\rm obs,G}<r_{\rm bar,G}$ (cuspy) (blue histogram)}
	\label{Fig: residuals}
\end{figure}

In Fig.~\ref{Fig: residuals} we show the distribution of normalized residuals $R(r_{j,G})$ from Eq.~\eqref{Eq:residuals} of the SPARC data after quality cuts and the $\delta v_{\rm obs}/v_{\rm obs}<0.1$ cut employed in \cite{McGaugh:2016leg,Lelli:2017vgz,Lelli:2016zqa} with respect to the MOND modified inertia prediction with $g_0=1.2\cdot 10^{-10} m/s^2$. The gray histograms, highlighted with black dots in the midle of the bins, show the residuals of all SPARC data points at large radii, i.e. with $r_{\rm j,G}>r_{\rm bar,G}$. These points are seen to follow the gaussian of unit width superimposed on the figure as expected of a good fit. 
The left panel also shows the residuals of data at small radii, meaning $r_{\rm j,G}\leq r_{\rm bar,G}$ from the (cored) galaxies with $r_{\rm obs,G}>r_{\rm bar,G}$ (red histogram). These residuals are not gaussian but skewed towards large negative residuals consistent with these points being below the MOND modified inertia prediction in general, as they are for our example galaxies. This is also what MOND modified gravity would in general predict they should be, however for our examples in Fig.~\ref{Fig: Galaxiescorecusp} the specific cores of MOND modified gravity do not match data well. In the right panel we also show the residuals of data at small radii $r_{\rm j,G} \leq r_{\rm bar,G}$ from (cuspy) galaxies with $r_{\rm obs,G}<r_{\rm bar,G}$ (blue histogram). These residuals are skewed towards large positive residuals consistent with these points lying above the MOND modified inertia predictions and therefore also above the MOND modified gravity predictions in $g2$-space.
We note that 7 of the galaxies in SPARC with steeply rising (cuspy) rotation curves are starburst dwarf galaxies where data may not represent the underlying gravitational potential faithfully as discussed in \cite{2016MNRAS.462.3628R,Santos-Santos:2018}. However some of those galaxies are eliminated by the data quality cut employed here and they are not among the galaxies presented in Fig.~\ref{Fig: Galaxiescorecusp}.

\newpage
\subsection{Model fits to data}

To reduce systematic uncertainties in the data before performing quantitative fits of MOND to the galaxies in Fig.~\ref{Fig: Galaxiescorecusp} we define the radius of maximum baryonic acceleration in the data (in analogy with the model $r_N$ above ) and acceleration ratios: 

\begin{equation}
 g_{\rm bar}(r_{\rm bar,G}) = {\rm max} \{ g_{\rm bar}(r_{j,G}) \} ; \quad   \hat{g}_{obs}(r_{j,G})\equiv\frac{g_{\rm obs}(r_{j,G})}{g_{obs}(r_{\rm bar,G})}, \quad\hat{g}_{bar}(r_{j,G})\equiv \frac{g_{bar}(r_{j,G})}{g_{bar}(r_{bar,G})},
\label{Eq:datamax1}
\end{equation}

\begin{figure}[htp!]
	\centering
	      \includegraphics[width=0.3\textwidth]{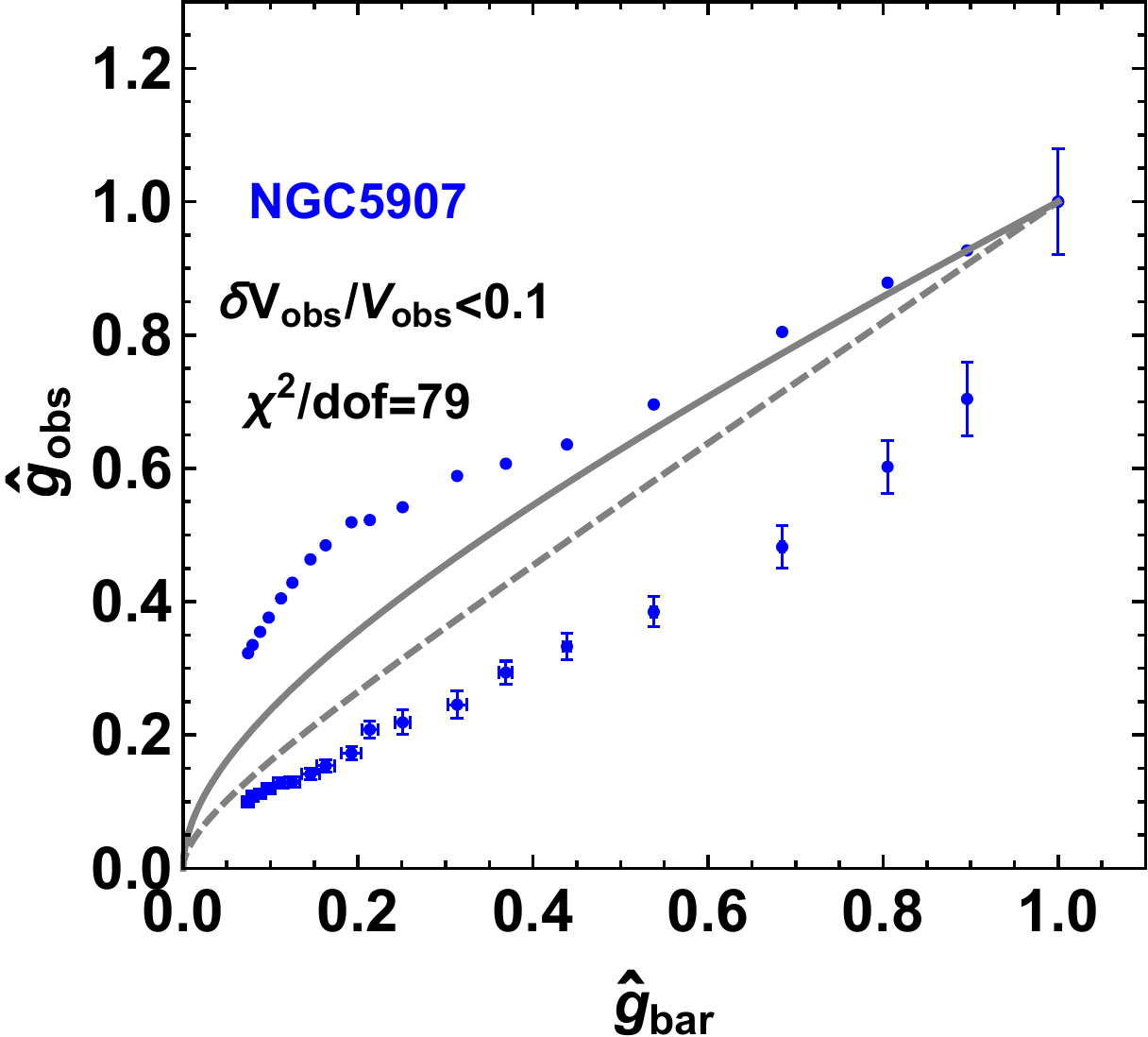}
	\includegraphics[width=0.3\textwidth]{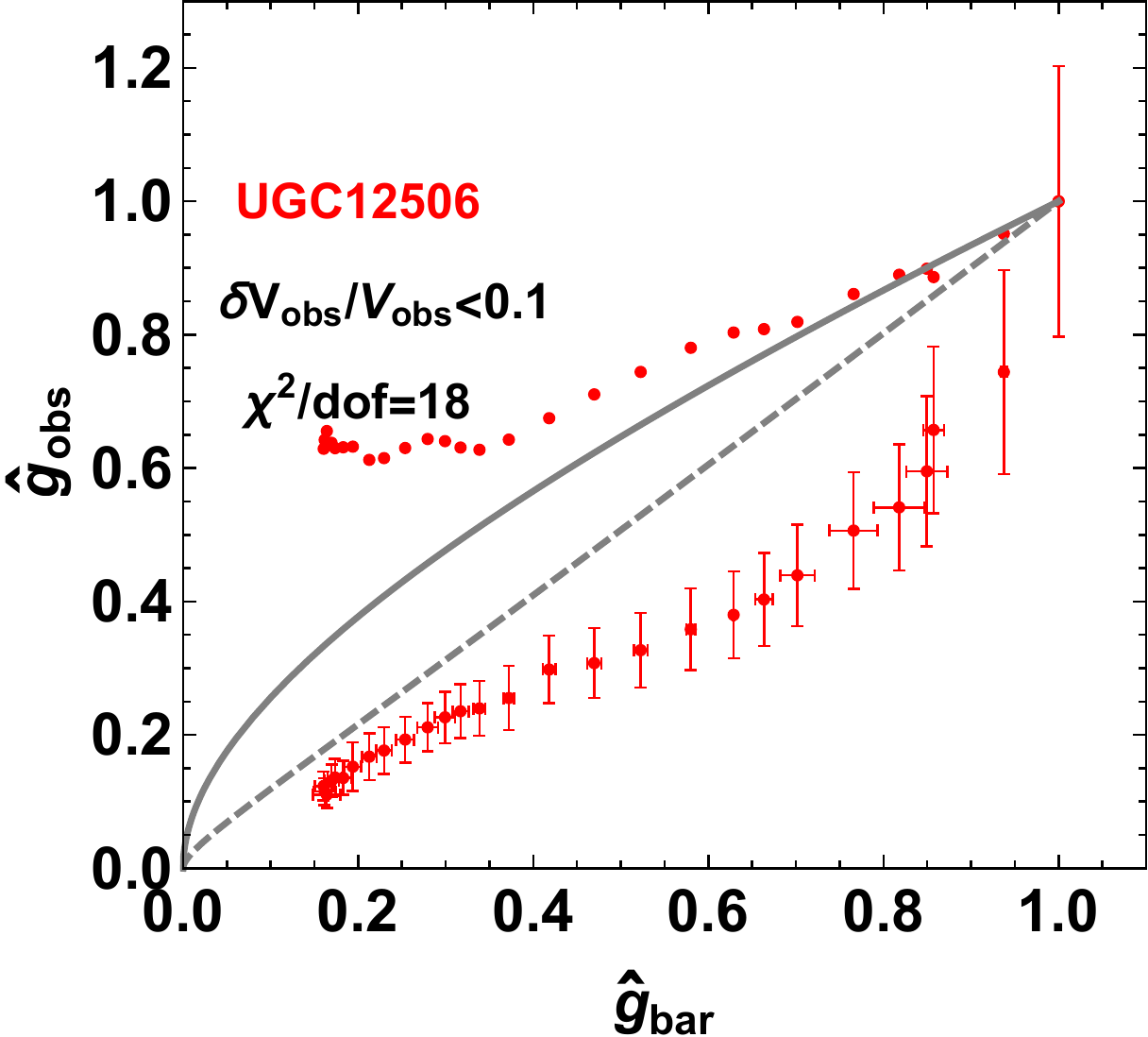}	
	                \includegraphics[width=0.3\textwidth]{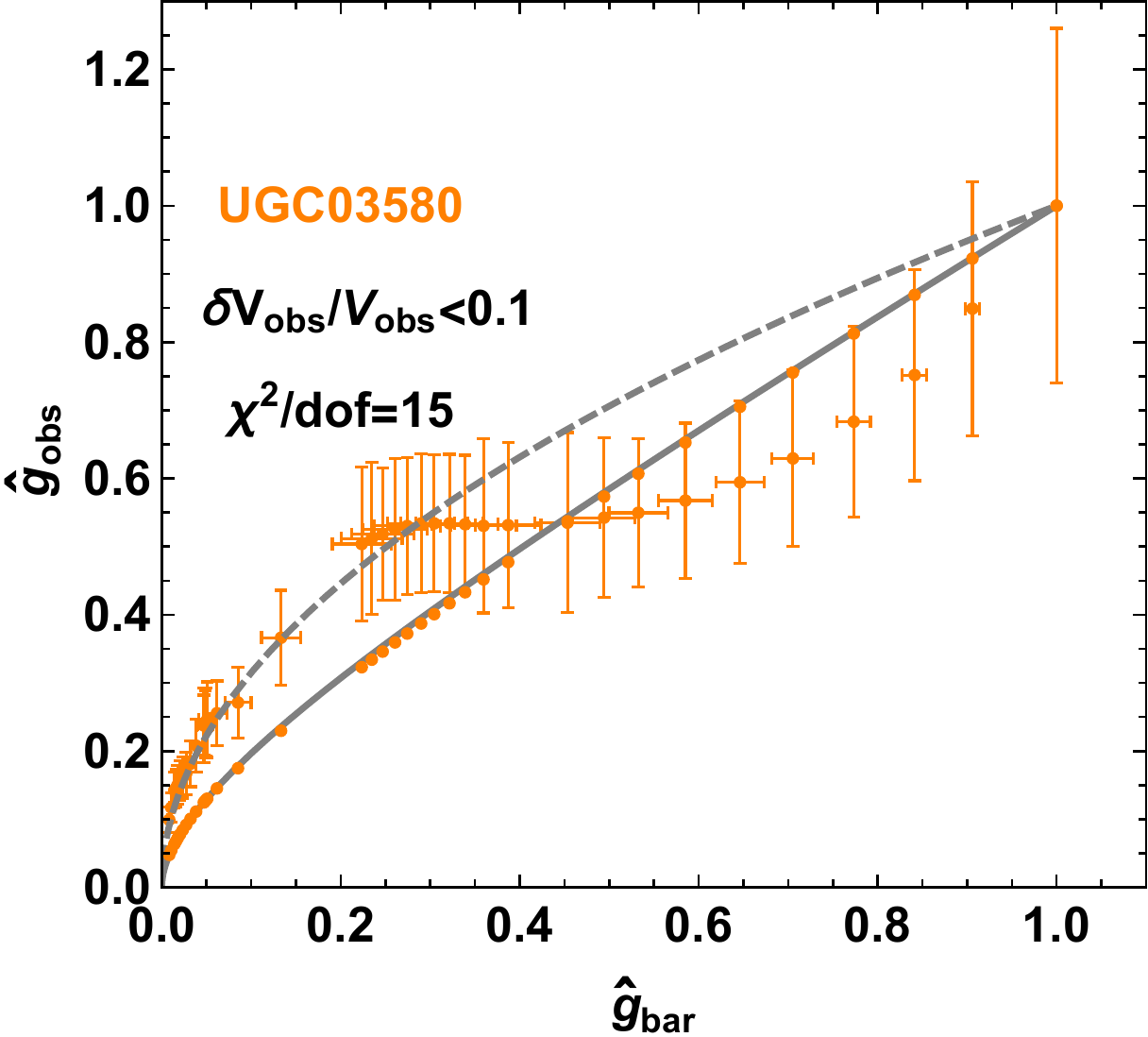}
	\caption
	{Acceleration ratio curves $(\hat{g}_{\rm bar} ,\hat{g}_{\rm obs} )$  (points with errors)  in $\hat{g}2$-space for 3 of the cuspy SPARC from the right hand panel of Fig.~\ref{Fig: Galaxiescorecusp}, compared to the MOND modified inertia curve with $g_0=1.2 \times 10^{-10}$ (gray solid), MOND modified inertia curve with best fit value of $g_{0,min}$ for each galaxy (gray dashed) and MOND modified gravity in the Brada-Milgrom approximation curves (points without errors). 
	Also shown are the $\chi^2$/dof values of the MOND modified inertia curve with $g_0=1.2 \times 10^{-10}$. 
		}
	\label{Fig:cuspyscaled}
\end{figure}

\begin{figure}[htp!]
	\centering
	\includegraphics[width=0.3\textwidth]{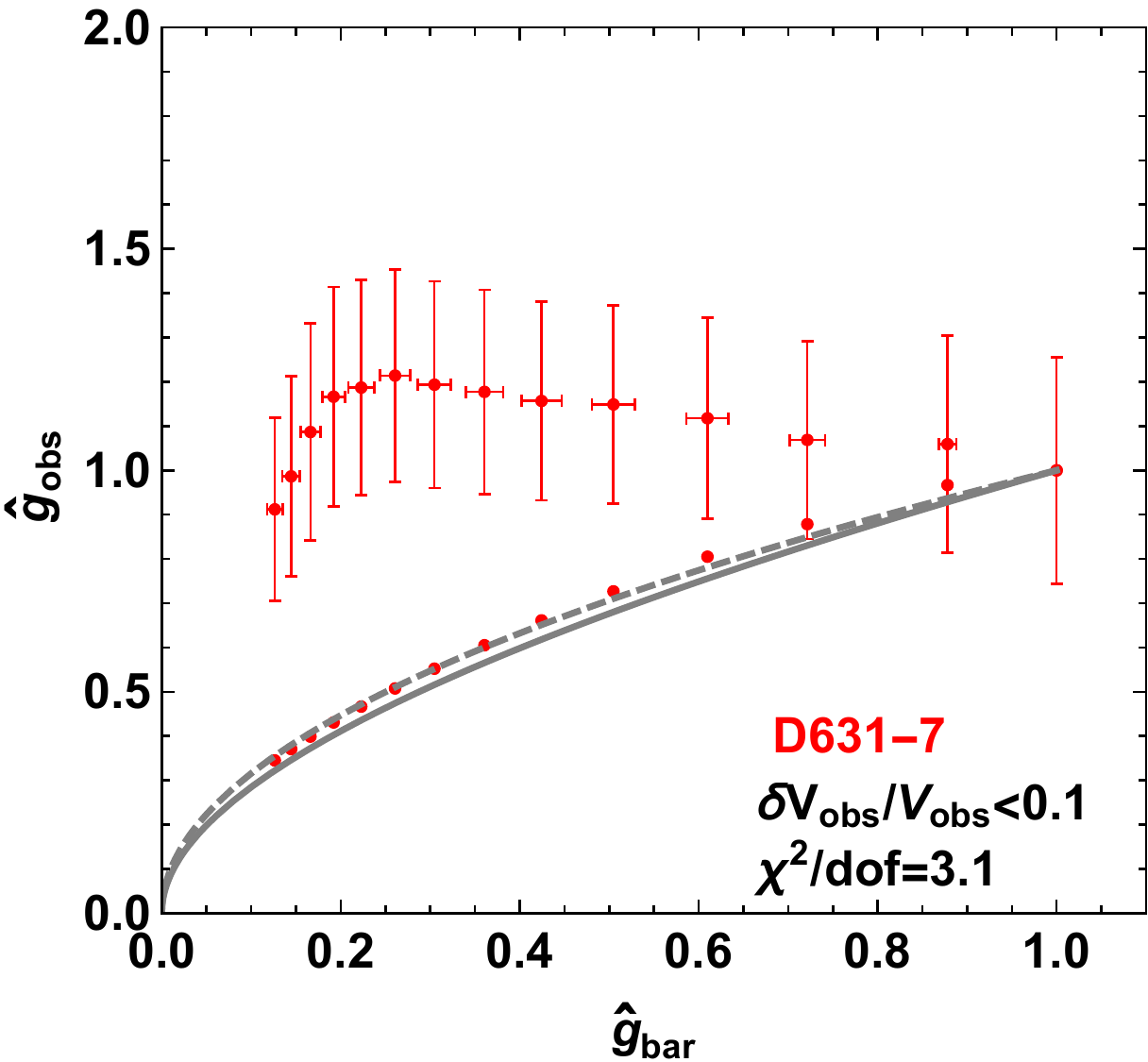}	
        \includegraphics[width=0.3\textwidth]{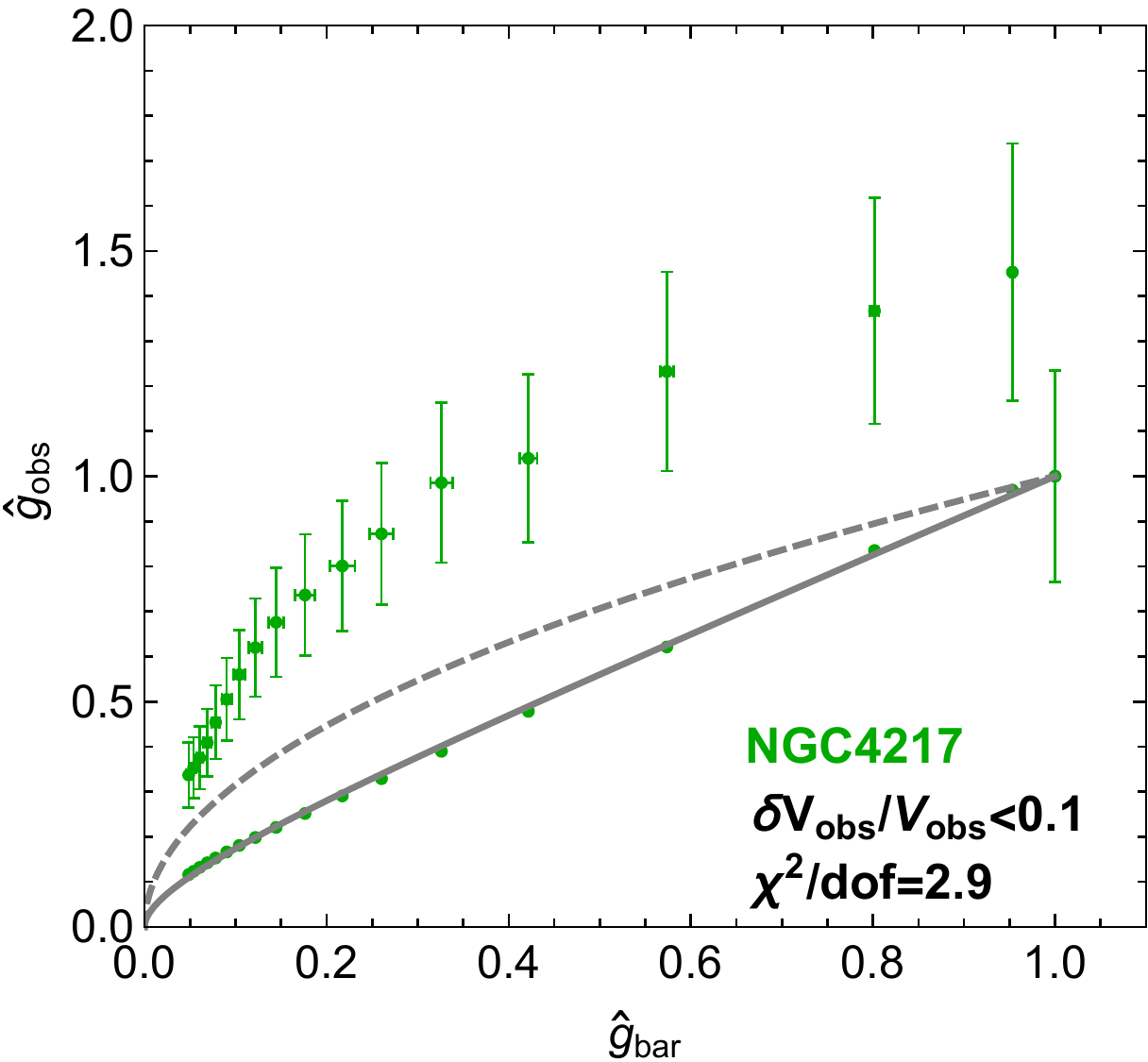}     
	                \includegraphics[width=0.3\textwidth]{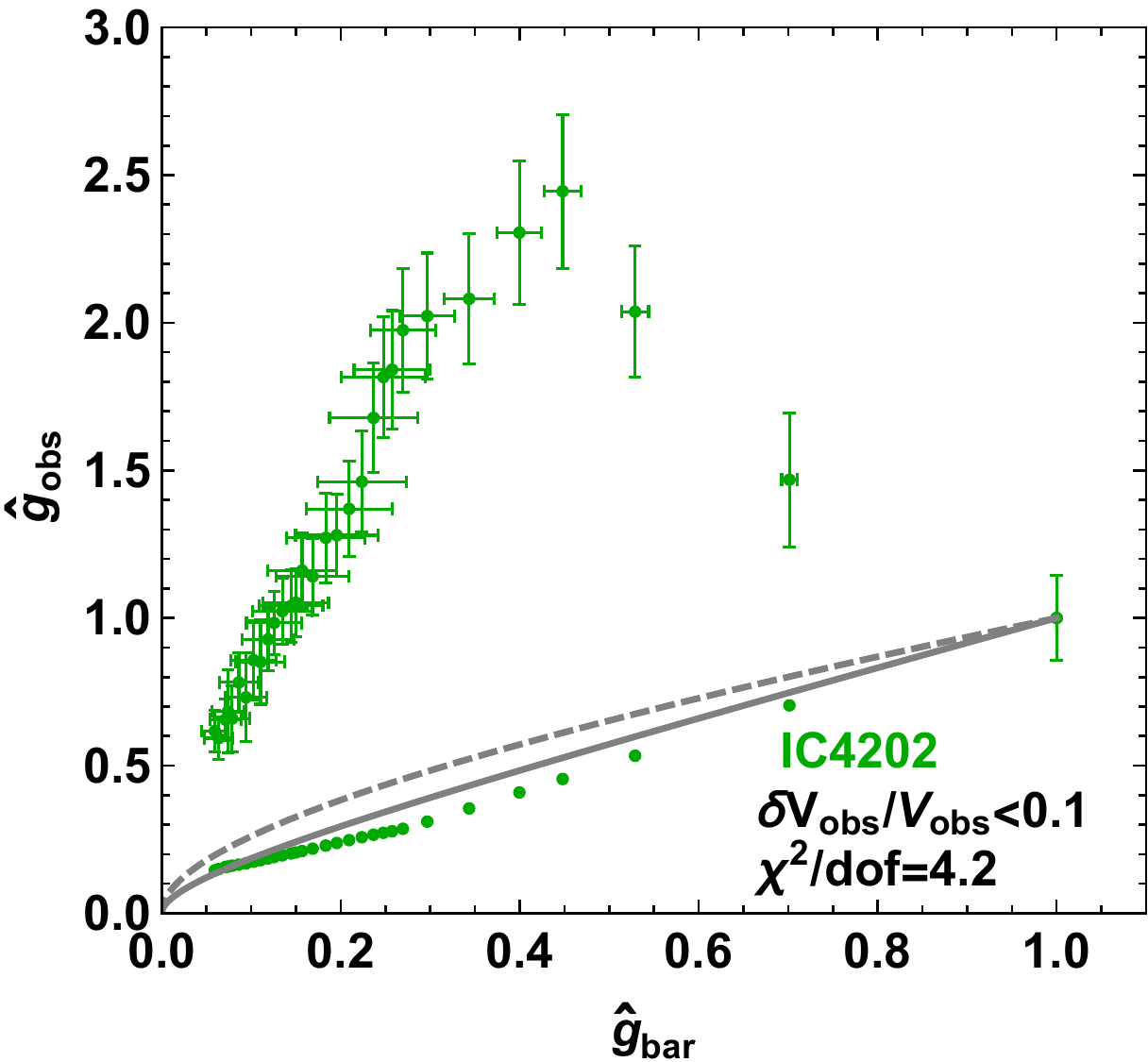}
	\caption
	{The $\hat{g}2$-space data points (points with errors) of 3 of the cored SPARC galaxies shown in the left hand panel of Fig.~\ref{Fig: Galaxiescorecusp}, compared to the MOND modified inertia curve with $g_0=1.2 \times 10^{-10}$ (gray solid), MOND modified inertia curve with best fit value of $g_0$ for each galaxy (gray dashed) and MOND modified gravity (Brada-Milgrom approximation) curves (points without errors). Note that most $\hat{g}_{\rm bar}$ uncertainties are small for most cases because the ratios are constructed to reduce systematic uncertainties. Also shown are the $\chi^2$/dof values for the dashed best fit MOND modified inertia curves.
		}
	\label{Fig:coredscaled}
\end{figure}

These ratios, introduced in \cite{Frandsen:2018ftj}, eliminate the systematic uncertainties on galactic distance and inclination angle for $g_{\rm obs}$ and they significantly reduce the systematic error from gas measurements and mass to light ratios for  $g_{\rm bar}$. 
The remaining uncertainty contributions to $\delta \hat{g}_{\rm bar} (r_{j,G})$ and $\delta \hat{g}_{\rm obs} (r_{j,G})$ following from Eq.~\eqref{Eq:accerrors} are
given in  \cite{Frandsen:2018ftj} and reproduced in appendix~\ref{App:Uncertainties}.
 Using the acceleration ratios we construct the $\chi^2_G$ of each galaxy compared to the model acceleration 
\begin{equation}
\chi^2_G = \sum_{j,j'} \left(\hat{g}_{\rm obs}  (r_{j,G})-\hat{g}_M (r_{j,G}) \right)  V^{-1}_{jj'}      \left(\hat{g}_{\rm obs}  (r_{j',G})-\hat{g}_M (r_{j',G}) \right)
\label{Eq:chisquare}
\end{equation}
where $\hat{g}_M (r_{j,G})= \hat{g}_M(\hat{g}_{\rm bar}  (r_{j,G}))$ is the MOND model prediction with the data point $\hat{g}_{\rm bar}  (r_{j,G})$ as input and the inverse variance matrix $V^{-1}_{jj'}$ given in in appendix~\ref{App:Uncertainties} takes into account the systematic uncertainty from the normalization point common to all acceleration ratios within a single galaxy.
We neglect the uncertainties in  $ \hat{g}_{\rm bar}(r_{\rm bar,G}) $ which are small compared to those in $ \hat{g}_{\rm obs} (r_{j,G})$ for most data points as seen in Fig~\ref{Fig:coredscaled} and Fig.~\ref{Fig:cuspyscaled} where the acceleration ratios with errors are shown.  

\bigskip
{\bf Cuspy SPARC galaxies:}
We first study 3 examples of the cuspy galaxies from the top right panel in Fig.~\ref{Fig: Galaxiescorecusp}. Their corresponding data curves with errors in the normalized $\hat{g}$ variables and after imposing the data cut $\delta v_{\rm obs}/v_{\rm obs}<0.1$ used in \cite{Lelli:2017vgz,Lelli:2016zqa} are shown in Fig.~\ref{Fig:cuspyscaled}. Also shown are the MOND modified inertia curves with $g_0=1.2 \times 10^{-10} m/s^2$ (solid gray) and with the best fit value $g_{0,min}$ that minimizes the $\chi_G^2$ in Eq.~\eqref{Eq:chisquare} (dashed gray). Finally the MOND modified gravity model curve using the Brada Milgrom approximation is shown as points without errors in same colour as data. 
The uncertainties on $\hat{g}_{\rm bar}$ are indeed small compared to those on $\hat{g}_{\rm obs}$ for most points.
We can therefore compute the $\chi^2_G$ value for each of these galaxies with respect to MOND using Eq.~\ref{Eq:chisquare}.
We give the $\chi^2_G$ value of MOND modified inertia with $g_0=1.2 \times 10^{-10} m/s^2$  fixed \cite{McGaugh:2016leg} in the second row of table~\ref{Table:fitscuspy} and the minimum $\chi^2_G$ with the corresponding$g_{0,min}$ value in the third row. 
As already discussed and as clear from the figures, the $\chi^2_G$ values of MOND modified gravity will be larger or equal to those of MOND modified inertia quoted. 
\begin{table}[htp!]
\centering
\begin{tabular}{ |p{2.5cm}||p{3cm}|p{4cm}|p{5cm}|p{5cm}| }
 \hline
Cuspy Galaxy & $\chi^2_G$/dof  ($g_{0}$) & $\chi^2_{G,min}$/dof ($g_{0,min}$)\\
 \hline
NGC5907 & $79  (1.2 \times 10^{-10})$& $14.5 \ (1.4 \times 10^{-11})$\\
UGC12506  & $18  (1.2 \times 10^{-10})$  &$1.8 \ (1.5 \times 10^{-12})$\\
UGC03580 & $15 (1.2 \times 10^{-10})$&  $2.9 (\infty)$\\
 \hline
\end{tabular}
\caption{Acceleration ratio curves $(\hat{g}_{\rm bar} ,\hat{g}_{\rm obs} )$  (points with errors)  in $\hat{g}2$-space for 3 of the cuspy SPARC from the right hand panel of Fig.~\ref{Fig: Galaxiescorecusp}, compared to the MOND modified inertia curve with $g_0=1.2 \times 10^{-10}$ (gray solid), MOND modified inertia curve with best fit value of $g_{0,min}$ for each galaxy (gray dashed) and MOND modified gravity in the Brada-Milgrom approximation curves (points without errors). 
}
\label{Table:fitscuspy}
\end{table}

\bigskip
{\bf Cored SPARC galaxies }
We next consider fits of MOND to three of the cored galaxies shown in $g2$-space in the top left panel of Fig.~\ref{Fig: Galaxiescorecusp}. The corresponding acceleration ratio curves are shown in Fig.~\ref{Fig:coredscaled}.
The $\chi^2_G$ values of these galaxies with respect to MOND modified inertia, analogous to those above for cuspy galaxies, are listed in table~\ref{Table:fitscored}.
\begin{table}[htp!]
\centering
\begin{tabular}{ |p{2.5cm}||p{3cm}|p{5cm}|p{3cm}|}
 \hline
Cored Galaxy & $\chi^2_G$/dof  ($g_{0}$) & $\chi^2_{G,min}$/dof ($g_{0,min}$)\\
 \hline
D631-7 &   $3.1 (1.2 \times 10^{-10})$ & $2.9 (\infty)$\\
NGC4217 &   $2.9 (1.2 \times 10^{-10})$  & $1 (\infty)$ \\
IC4202&   $4.2 (1.2 \times 10^{-10})$  & $4 (1.3\times 10^{-9})$ \\
 \hline
\end{tabular}
\caption{$\chi^2$ values for fits of MOND modified inertia to cored galaxy examples. $\chi^2$ values with $g_0=1.2 \times 10^{-10} m/s^2$ fixed are shown in second column and $\chi^2$ values for the best fit $g_0$ values are shown in the third column. As discussed in the text the negligible difference between MOND modified inertia and MOND modified gravity for these galaxies imply that the chi-square values for MOND modified gravity are the same to a good approximation. 
}
\label{Table:fitscored}
\end{table}
Again we find large $\chi^2_G$ values for the fits to MOND modified inertia and given the very little difference between this and MOND modified gravity seen on the figures we can take these values also for MOND modified gravity.

\newpage

\section{summary}
In this paper we have employed a new definition of cuspy and cored galactic acceleration curve geometries, that is applicable to DM and modified gravity models of the missing mass.  
Cuspy and cored curves are defined relative to the curves of MOND modified inertia which we take as a neutral reference and the classification in the space of baryonic and total accelerations ($g2$-space) is that proposed in \cite{Frandsen:2018ftj}. It is summarized in table~\ref{Table:geometries} and illustrated in Fig~\ref{Fig:geometries}.

Based on this we have elucidated a cusp-core challenge for Modified Newtonian Dynamics which is distinct from the cusp-core problem discussed in the context of DM: MOND modified gravity leads to cored rotation curves for isolated galaxies as a consequence of the solenoidal field $\vec{S}$ in Eq.~\eqref{MONDgrav}. 
The cored curves from MOND modified gravity models are most pronounced for galaxies with the most disk like baryonic matter distribution and in the limit of spherical distributions they reduce to
the neutral curves of MOND modified inertia with the functional relation $g_{\rm tot}=g_{\rm tot}(g_N)$. This is shown in Fig.~\ref{Fig:MONDcurves}, and discussed in section~\ref{Sec: geometries}. The cored and neutral curves of MOND may be contrasted with e.g. the cuspy curves from NFW DM density profiles - arising in N-body simulations of DM structure formation without baryons.

Examples of cored and cuspy acceleration curves in $g2$-space, and corresponding rotation speed curves, from the {\tt SPARC} data base are shown in the top and bottom panels of Fig.~\ref{Fig: Galaxiescorecusp}. 
We fit MOND modified inertia and MOND modified gravity to some of these galaxies in Fig.~\ref{Fig:cuspyscaled} and Fig.~\ref{Fig:coredscaled} with the acceleration scale $g_0$ as fit parameter. To eliminate systematic uncertainties from galaxy distance and inclination angle and reduce the systematic uncertainty from mass to light ratios the fit is performed on ratios of accelerations $\hat{g}_{\rm bar}$ and $\hat{g}_{\rm obs}$ defined in Eq.~\eqref{Eq:datamax1}.
As summarized in table~\ref{Table:fitscuspy} and table~\ref{Table:fitscored} the fits return large $\chi^2$ values and significant variation of the best fit values of $g_0$ between different galaxies. Also these best fit values of $g_0$ deviate considerably from the value  $g_0=1.2 \times 10^{-10} m/s^2$ which was found in \cite{Lelli:2017vgz} as a best fit to the entire SPARC data set. These deviations are  visually clear from Fig.~\ref{Fig: Galaxiescorecusp}
and is in line with our previous findings that the neutral geometry of MOND modified inertia, specifically the prediction $g_{\rm tot}(r_{\rm tot})=g_{\rm tot}(r_{\rm N})$ is in tension with the full SPARC data set of ca. 150 galaxies \cite{Frandsen:2018ftj}, independent of the interpolation function.

Baryonic feedback from supernovae can change cuspy DM profiles into cored ones in some cases \cite{Read:2004xc,Teyssier:2012ie,DiCintio:2013qxa}. In the future, it would be interesting to investigate whether e.g. the external field effect in MOND or a radial dependence of the mass-to-light ratios impact the cusp-core challenge for MOND.

\bigskip
{\bf Acknowledgments:}
We thank J.Read, W.-C. Huang and J. Smirnov for discussions comments on the draft. 
The authors acknowledge partial funding from The Council For Independent Research, grant number DFF 6108-00623. The CP3-Origins center is partially funded by the Danish National Research Foundation, grant number DNRF90.

\newpage

\appendix

\section{3MN Model}
As discussed in the main text, the sum of 3 Miyamoto-Nagai disks can be used to model an exponential disk profile using the following procedure \cite{3MN}. 
First the single scale lenght used in all three potentials $b$ is found in terms of the exponential disk scale length $R_d$ and scale height $h_z$ as 
	\begin{gather}
			\frac{b}{R_d}=-0.269\left( \frac{h_z}{R_d} \right)^3+1.080\left( \frac{h_z}{R_d} \right)^2+1.092 \left( \frac{h_z}{R_d} \right),\\
		\end{gather}
The remaining 6 parameters, the 3 mass scale parameters $M_{MN,1}$, $M_{MN,2}$, $M_{MN,3}$ and the three scale height parameters $a_1$, $a_2$, $a_3$ are found from the equation
\begin{gather}
			\text{parameter}=k_1 \left( \frac{b}{R_d}\right)^4+k_2\left( \frac{b}{R_d}\right)^3+k_3\left( \frac{b}{R_d}\right)^2+k_4\left( \frac{b}{R_d}\right)+k_5.
		\end{gather}
with the numerical parameters $k_i$ given by 
		\begin{table}[]
			\begin{tabular}{llllll}
				Parameter      & $k_1$     & $k_2$     & $k_3$     & $k_4$     & $k_5$     \\ \hline
				$M_{MN,1}/M_d$ & $-0.0090$ & $0.0640$  & $-0.1653$ & $0.1164$  & $1.9487$  \\
				$M_{MN,2}/M_d$ & $0.0173$  & $-0.0903$ & $0.0877$  & $0.2029$  & $-1.3077$ \\
				$M_{MN,3}/M_d$ & $-0.0051$ & $0.0287$  & $-0.0361$ & $-0.0544$ & $0.2242$  \\
				$a_1/R_d$      & $-0.0358$ & $0.2610$  & $-0.6987$ & $-0.1193$ & $2.0074$  \\
				$a_2/R_d$      & $-0.0830$ & $0.4992$  & $-0.7967$ & $-1.2966$ & $4.4441$  \\
				$a_3/R_d$      & $-0.0247$ & $0.1718$  & $-0.4124$ & $-0.5944$ & $0.7333$ 
			\end{tabular}
			\caption{The parameters used for the 3MN model. Table is taken from \cite{3MN}}
			\label{Table 3MN}
		\end{table}
The 3MN models found by this table matches the analytical exponential disk $<1.0 \%$ out to $4 R_d$ and $<3.3 \%$ out to $10 R_d$ \cite{3MN}. 
For our purposes the 3MN model is convenient because we can use it to interpolate explicitly between spherical and disk-like matter distributions. 

\section{Uncertainties on acceleration ratios and variance matrix}
\label{App:Uncertainties}
 The uncertainty contributions to $\delta \hat{g}_{\rm bar} (r_{j,G})$ and $\delta \hat{g}_{\rm obs} (r_{j,G})$ following from Eq.~\eqref{Eq:accerrors} are
\begin{equation}
\begin{split}
\delta \hat{g}_{\rm obs} (r_{j,G}) &=  \sqrt{\delta_1 \hat{g}_{\rm obs} (r_{j,G})^2 + \delta_{2} \hat{g}_{\rm obs} (r_{j,G})^2}
\\
\delta \hat{g}_{\rm bar}  (r_{j,G})&= \hat{g}_{\rm bar}(r_{j,G}) \sqrt{
	\left( \Delta g_{gas} (r_{j,G}) \right)^2  + \left( \Delta \Upsilon (r_{j,G}) \right)^2} , \\
\delta_1 \hat{g}_{\rm obs} (r_{j,G}) &= \hat{g}_{\rm obs}  (r_{j,G}) \left(\frac{2 \delta v_{\rm obs} (r_{j,G})}{ v_{\rm obs} (r_{j,G})}\right) \\
 \delta_{2} \hat{g}_{\rm obs} (r_{j,G}) &= \hat{g}_{\rm obs}  (r_{j,G})  \left(\frac{2 \delta v_{ \rm obs }(r_{\rm bar,G }) }{ v_{ \rm obs }(r_{\rm bar,G })}\right)   \\
\Delta g_{gas} (r_{j,G})&=  \left( \frac{\delta g_{\rm gas}(r_{j,G})}{g_{\rm bar}(r_{j,G})} - \frac{ \delta g_{\rm gas}(r_{\rm bar,G})}{g_{\rm bar }(r_{\rm bar,G })} \right) \\
\Delta \Upsilon (r_{j,G})& = \sum_{k={\rm disk, bulge}} \delta \Upsilon_k  \left( \frac{v_k^2(r_{j,G})}{v_{\rm bar}^2(r_{j,G})} - \frac{ v_{k}^2(r_{\rm bar,G })}{v_{\rm bar }^2(r_{\rm bar,G })} \right) 
\end{split}
\label{Eq:haterrors1}
\end{equation}
where we have separated the uncertainty contribution $\delta_1 \hat{g}_{\rm obs} (r_{j,G})$ which is random for all points within a galaxy and the contribution from the normalization $ \delta_{2} \hat{g}_{\rm obs} (r_{j,G}) $ which is a systematic for all data points within a single galaxy. 
With these ratios we construct the $\chi^2$ of each galaxy compared to the model acceleration 
\begin{equation}
\chi^2_G = \sum_{j,j'} \left(\hat{g}_{\rm obs}  (r_{j,G})-\hat{g}_M (r_{j,G}) \right)  V^{-1}_{jj'}      \left(\hat{g}_{\rm obs}  (r_{j,G})-\hat{g}_M (r_{j,G}) \right)
\label{Eq:chisquareapp}
\end{equation}
where $\hat{g}_M (r_{j,G})= \hat{g}_M(\hat{g}_{\rm bar}  (r_{j,G}))$ is the MOND model prediction with the data point $\hat{g}_{\rm bar}  (r_{j,G}$ as input and the inverse variance matrix, e.g. \cite{Stump:2001gu} is 
\begin{equation}
V^{-1}_{jj'} =   \frac{\delta_{jj'}}{\delta_1 \hat{g}_{\rm obs} (r_{j,G})^2 }- \frac{\delta_2 \hat{g}_{\rm obs} (r_{j,G})\delta_2 \hat{g}_{\rm obs} (r_{j',G})}{\delta_1 \hat{g}_{\rm obs} (r_{j,G})^2 \delta_1 \hat{g}_{\rm obs} (r_{j',G})^2 } A^{-1} , \quad A^{-1} = 1+\sum_j  \frac{\delta_2 \hat{g}_{\rm obs} (r_{j,G})^2}{\delta_1 \hat{g}_{\rm obs} (r_{j,G})^2  }
\end{equation}
The second term in $V^{-1}_{jj'} $ comes from the systematic uncertainties $ \delta_2 \hat{g}_{\rm obs} (r_{j,G})$ introduced due to the common normalization point. We neglect the uncertainties in  $ \hat{g}_{\rm bar}(r_{\rm bar,G}) $ which are small compared to those in $ \hat{g}_{\rm obs} (r_{j,G})$ for most data points as seen in Fig~\ref{Fig:coredscaled} and Fig.~\ref{Fig:cuspyscaled}.

  %%%%%%%%%%%%%%%%%%%%%%%%%%%%%%%%%%%%%  
\bibliography{rot}
\bibliographystyle{hunsrt}
  %%%%%%%%%%%%%%%%%%%%%%%%%%%%%%%%%%%%%  

\end{document}